\pdfoutput=1
\newcommand*{\ATLASLATEXPATH}{latex/}
\documentclass[cernpreprint, UKenglish, texlive=2016, texmf]{atlasdoc}
\usepackage{\ATLASLATEXPATH atlaspackage}
\usepackage{\ATLASLATEXPATH atlasbiblatex}
\usepackage{\ATLASLATEXPATH atlascontribute}
\usepackage{\ATLASLATEXPATH atlasphysics}

\graphicspath{{logos/}{figures/}}

\usepackage{lscape}

\AtlasTitle{Measurement of the production cross section of three isolated photons in \pp collisions at \rts = 8 TeV using the ATLAS detector}

\author{The ATLAS Collaboration}

\AtlasJournalRef{Phys. Lett. B 781 (2018) 55}
\AtlasDOI{DOI:10.1016/j.physletb.2018.03.057}

\AtlasAbstract{%
  A measurement of the production of three isolated photons in proton--proton collisions at a centre-of-mass energy \rts = 8 TeV is reported. The results are based on an integrated luminosity of 20.2 \ifb collected with the ATLAS detector at the LHC. The differential cross sections are measured as functions of the transverse energy of each photon, the difference in azimuthal angle and in pseudorapidity between pairs of photons, the invariant mass of pairs of photons, and the invariant mass of the triphoton system. A measurement of the inclusive fiducial cross section is also reported. Next-to-leading-order perturbative QCD predictions are compared to the cross-section measurements. The predictions underestimate the measurement of the inclusive fiducial cross section and the differential measurements at low photon transverse energies and invariant masses. They provide adequate descriptions of the measurements at high values of the photon transverse energies, invariant mass of pairs of photons, and invariant mass of the triphoton system.
}

\PreprintIdNumber{CERN-EP-2017-302} 

\hypersetup{pdftitle={ATLAS document},pdfauthor={The ATLAS Collaboration}}

\begin{document}

\maketitle

\section{Introduction}
\label{sec:intro}

The production of three prompt photons in proton--proton ($pp$) collisions, $pp\rightarrow \gamma\gamma\gamma$ + $X$, provides a testing ground for perturbative quantum chromodynamics (pQCD). This process is rare in the Standard Model (SM) since the leading-order (LO) contribution to triphoton production is of order $\alpha_{\text{EM}}^{3}$. The measurement of triphoton production can be performed in a broader range of 
kinematic regions than in 2~$\rightarrow$~2 reactions such as inclusive-photon~\cite{STDM-2011-04,STDM-2012-16,STDM-2014-09,CMS-QCD-10-019} and diphoton~\cite{STDM-2012-05,STDM-2015-15,CMS-SMP-13-001} production. This provides a complementary test of pQCD in processes with photons in the final state.

Precise measurements of triphoton production can be used to improve the description of this process in Monte Carlo (MC) models. In addition, SM triphoton production provides one of the main irreducible backgrounds for some beyond-the-SM (BSM) searches. Potential BSM processes include the associated production of a photon and an exotic neutral particle decaying into a photon pair ($q\overline{q}\rightarrow X^{0}\gamma$), where $X^{0}$ can be a Kaluza--Klein graviton ($G_{\text{KK}}$)~\cite{physlettB:429:1998,physrevlett:83:1999,physrevD:92:2015} or a pseudoscalar ($a$)~\cite{physrevD:86:2012}. Moreover, triphoton production is also the main background to the predicted decay of the $Z$ boson into three photons. The current upper limit at 95\% confidence level on the branching fraction for $Z\rightarrow3\gamma$ is 2.2~$\times$~10$^{-6}$~\cite{EXOT-2013-24}.

Three photons can be produced via two main mechanisms: direct and fragmentation production. In the case of the direct production 
process, three photons are produced in the hard interaction via the annihilation of an initial-state quark--antiquark pair ($q\overline{q}\rightarrow\gamma\gamma\gamma$). In the fragmentation process, at least one of the photons arises from the fragmentation of a high-transverse-momentum (high-$p_{\text T}$) parton ($qg\rightarrow\gamma\gamma q[\gamma]$). Direct photons are typically isolated, while those originating from the fragmentation process are usually accompanied by nearby partons.~Measurements of final-state photons include an isolation requirement to reduce background contributions from neutral-hadron decays into photons. As a consequence, signal processes with one or more fragmentation photons are also suppressed.

This Letter presents measurements of three-photon production. The analysis is performed using 20.2$\pm$0.4~fb$^{-1}$ of ATLAS data at a centre-of-mass energy of $\sqrt{s}$~=~8~TeV~\cite{DAPR-2013-01}.~The measurements study the topology and kinematics of the individual photons, pairs of photons, and the three-photon system. Differential cross sections are measured as functions of the transverse energy\footnote{ATLAS uses a right-handed coordinate system with its origin at the nominal interaction point (IP) in the centre of the detector and the $z$-axis along the beam pipe. The $x$-axis points from the IP to the centre of the LHC ring,and the $y$-axis points upwards. Cylindrical coordinates ($r$,$\phi$) are used in the transverse plane, $\phi$ being the azimuthal angle around the $z$-axis measured in radians. The pseudorapidity is defined in terms of the polar angle $\theta$ as $\eta$ = - ln tan($\theta$/2).~The transverse energy is defined as $E_{\text T}$  = $E$  sin$\theta$, where $E$ is the energy.} of the leading photon ($E_{\text T}^{\gamma_{1}}$), the second-highest-$E_{\text T}$ photon ($E_{\text T}^{\gamma_{2}}$) and the third-highest-$E_{\text T}$ photon ($E_{\text T}^{\gamma_{3}}$); the difference in azimuthal angle and in pseudorapidity between pairs of photons ($\Delta\phi^{\gamma_{1}\gamma_{2}}$, $\Delta\phi^{\gamma_{1}\gamma_{3}}$, $\Delta\phi^{\gamma_{2}\gamma_{3}}$, $|\Delta\eta^{\gamma_{1}\gamma_{2}}|$, $|\Delta\eta^{\gamma_{1}\gamma_{3}}|$, $|\Delta\eta^{\gamma_{2}\gamma_{3}}|$); the invariant mass of pairs of photons ($m^{\gamma_{1}\gamma_{2}}$, $m^{\gamma_{1}\gamma_{3}}$ and $m^{\gamma_{2}\gamma_{3}}$); and the invariant mass of the triphoton system ($m^{\gamma\gamma\gamma}$). A measurement of the inclusive fiducial cross section is also reported. Photons are required to be isolated based on the amount of transverse energy, excluding the photon contribution, inside a cone of size $\Delta R\equiv\sqrt{(\eta - \eta^{\gamma})^{2} + (\phi - \phi^{\gamma})^{2}}$~=~0.4 centred around each photon direction (defined by the photon pseudorapidity $\eta^{\gamma}$ and azimuthal angle $\phi^{\gamma}$). Finally, the measurements are compared to next-to-leading-order (NLO) QCD calculations.

\section{ATLAS detector}
\label{sec:detector}

The ATLAS detector~\cite{PERF-2007-01} is a multi-purpose detector
with a forward-backward symmetric cylindrical geometry. The
most relevant systems for the present measurement are
the inner detector, immersed in a 2 T magnetic field
produced by a thin superconducting solenoid, and the
calorimeters.~At small radii, the inner detector is made up of fine-granularity pixel 
and microstrip detectors. These silicon-based detectors cover the pseudorapidity
range $|\eta|<2.5$. A gas-filled straw-tube transition radiation
tracker complements the silicon tracker at larger radii in the 
range $|\eta|<2.0$ and also provides electron identification capabilities 
based on transition radiation. The electromagnetic calorimeter is a 
lead/liquid-argon sampling calorimeter with accordion geometry.  The calorimeter is
divided into a barrel section covering $|\eta|<1.475$ and two end-cap
sections covering $1.375<|\eta|<3.2$. For $|\eta|<2.5$ it is divided
into three layers in depth, which are finely segmented in $\eta$ and
$\phi$. A thin presampler layer, covering $|\eta|<1.8$, is used to
correct for fluctuations in upstream energy losses. The hadronic
calorimeter in the region $|\eta|<1.7$ uses steel absorbers with
scintillator tiles as the active medium. Liquid-argon with copper 
absorbers is used in the hadronic end-cap calorimeters, which cover the region $1.5<|\eta|<3.2$. 
Events are selected using a first-level trigger implemented in custom
electronics, which reduces the event rate to a value of 75~kHz
using a subset of detector information. Software algorithms with
access to the full detector information are then used in the
high-level trigger to yield a recorded event rate of about
400~Hz~\cite{PERF-2011-02}.
\section{Monte Carlo simulations and theoretical predictions}
\label{sec:mc_simulations}

\subsection{Monte Carlo simulations}
\label{sec:mc_description}
 The MC samples were generated to study the characteristics of the signal and background events. The MC program {\textsc MadGraph 5.1.4.4 }~\cite{jhep06:2011:128} 
interfaced with {\textsc Pythia 8.186}~\cite{cpc:178:852} was used to simulate signal events. The partonic subprocess was simulated by {\textsc MadGraph} 
to include the leading-order matrix element ($q\overline{q}\rightarrow\gamma\gamma\gamma$), whereas {\textsc Pythia} was added to include the initial- and final-state 
parton showers and the fragmentation of partons into hadrons.~The LO CTEQ6L1 parton distribution functions (PDFs)~\cite{jhep:0207:012} are used to parameterise the parton momentum distributions in the proton. To study the effect of the contribution of photon fragmentation, a {\textsc Pythia} MC sample supplemented by 
QED final-state radiation was generated with LO CTEQ6L1 PDFs. This sample includes the LO diphoton, photon+jet and dijet processes with 
initial-state and final-state radiation modelled by the parton shower (PS). 

The MC program {\textsc Sherpa 1.4.1}~\cite{jhep:0902:007} was used to estimate the background arising from electrons misreconstructed as photons. Three processes 
were simulated with at least one high-$p_{\text T}$ electron and photon in the final state:~$e^{+}e^{-}\gamma$, $e^{+}e^{-}\gamma\gamma$, and 
$e^{\pm}\nu_{e}\gamma\gamma$. The matrix elements were calculated with 
up to three final-state partons at LO in pQCD and used the CT10 PDFs at NLO~\cite{PhysRevD:82:2010}. The matrix elements were merged with the {\textsc Sherpa} parton-shower algorithm~\cite{jhep:03:038} following 
the ME+PS@LO prescription~\cite{jhep:05:053}. 

The generated signal and background event samples were passed through the {\textsc Geant4}-based~\cite{nim:a506:250} ATLAS detector and trigger simulation programs~\cite{epj:c70:823}. The signal and background samples include a simulation of the underlying event (UE) where {\textsc Pythia} event-generator parameters were set according to the ``AU2'' tune~\cite{ATL-PHYS-PUB-2012-003}. The generation of the simulated event samples includes the effect of multiple $pp$ interactions per bunch crossing, as well as the effect of the detector response to interactions from bunch crossings before or after the one containing the hard interaction. These MC events were weighted to reproduce the distribution of the average number of interactions per bunch crossing observed in the data. The generated MC events are reconstructed and analysed with the same program chain as the data. 

\subsection{Next-to-leading-order pQCD predictions}
\label{sec:nlo_description}

The NLO pQCD predictions presented in this Letter are computed using the programs {\textsc MCFM}~\cite{MCFMcite2,FERMILAB-PUB-14-033-T} and 
{{\textsc MadGraph5}\_aMC@NLO} 2.3.3~\cite{jhep:1407:079}. The strong coupling constant is calculated at two loops with $\alpha_{\text S}(m_{Z})$~=~0.118 
and the electromagnetic coupling constant is set to $\alpha_{\text{EM}}$~=~1/137. In addition, the number of massless quark flavours is set to five and the CT10 parameterisations of the proton PDFs at NLO are used.

The {\textsc MCFM} program includes NLO pQCD calculations of the direct contribution, whereas the 
production of a photon via parton fragmentation is estimated from the LO QCD matrix element multiplied by the BFG~II parton-to-photon fragmentation functions~\cite{EurPhysJ:C2:1998}. The renormalisation scale $\mu_{\text R}$, factorisation scale $\mu_{\text F}$ and fragmentation scale $\mu_{\text f}$ are chosen to be  $\mu_{\text R}$ = $\mu_{\text F}$ = $\mu_{\text f}$ = $m^{\gamma\gamma\gamma}$. In addition, the {\textsc MCFM} calculations are performed using an isolation criterion which requires the total transverse energy from the partons inside a cone of size $\Delta R$ = 0.4 around the photon direction to satisfy $E_{\text T}^{\text {iso}} <$ 10 GeV. The MCFM NLO pQCD predictions refer to the parton level while the measurements are performed at the particle level. Since the $E_{\text T}^{\text{iso}}$ requirement at the particle level is applied after the subtraction of the UE transverse energy, it is expected that parton-to-hadron corrections to the NLO pQCD predictions are small. This is confirmed by computing the ratio of the particle-level cross section for a {\textsc MadGraph} sample interfaced with {\textsc Pythia} with UE effects to the computed cross section without hadronisation and UE effects. The ratio is consistent with unity over the measured range of the variables under study. Therefore, no correction is applied to the MCFM NLO pQCD calculations. Deviations from unity of $O(1\%)$ on the parton-to-hadron correction factors are found when the hadronisation and UE effects are included using {\textsc Herwig++~7.0.1}~\cite{Bahr:2008pv}. Predictions based on other proton PDF sets, namely MSTW2008~\cite{mstw:2009epjc} and NNPDF2.1~\cite{nnpdf:2011npb}, are also computed. Differences of +5\% and +6\% in the calculation of the inclusive fiducial cross section are found using the MSTW2008 and NNPDF2.1 PDF sets, respectively, whereas the dependence of the shape of the differential cross sections on the PDF sets is found to be small.

{{\textsc MadGraph5}\_aMC@NLO} calculations include the NLO pQCD contribution of direct processes and apply a smoothly varying isolation cone to the photons~\cite{plb:429:369}. This isolation requirement regularises the photon collinear divergences which appear in the calculation of the matrix element and removes the contribution of photons resulting from the fragmentation of a parton: $E_{\text T}^{\text{iso}}(\Delta R) < E_{\text T}^{\gamma} (1-\cos~\Delta R)/(1-\cos~R_{0})$, where $R_{0}$ = 0.4 and $E_{\text T}^{\text{iso}}(\Delta R)$ is the sum of the transverse energies of the particles around the photon up to $\Delta R$. The {{\textsc MadGraph5}\_aMC@NLO} calculations are interfaced with {\textsc Pythia 8.212}~\cite{cpc:191:159} in the NLO+PS prescription to include the initial- and final-state parton showers and the hadronisation~\cite{jhep:0206029}.~The renormalisation and factorisation scales are chosen to be equal to the transverse mass of the clustered jets from the final state partons and photons defined in the matrix element. This choice follows the recommendations in Ref.~\cite{jhep:1407:079} when interfacing the {{\textsc MadGraph5}\_aMC@NLO} calculations to 
{\textsc Pythia}.~After the generation, the isolation value of the photon is computed by summing the transverse energy of all final-state 
particles (excluding muons and neutrinos) inside a cone of size $\Delta R$ = 0.4 around the photon candidate. Events with $E_{\text T}^{\text{iso}} >$~10~GeV for any of the photons are excluded. 
\section{Event selection}
\label{sec:data_selection}

The data considered in this analysis were taken in stable beam conditions and satisfy detector and data-quality requirements. Events are recorded using a diphoton trigger with a transverse energy threshold of 20 GeV. The trigger efficiency for pairs of isolated photons with 
$E_{\text T}^{\gamma}>$ 22~GeV and $|\eta^{\gamma}|<$ 2.37 is higher than 99\%. Events are required to have a reconstructed 
primary vertex with at least two associated tracks with $p_{\text T}~>$~500 MeV and $|\eta|<$~2.5, consistent with originating from the same 
three-dimensional spot within the luminous region of the colliding proton beams. If multiple primary vertices are reconstructed, 
the one with the highest sum of the $p_{\text T}^{2}$ of the associated tracks is selected as the primary vertex.

Photon and electron candidates are reconstructed from clusters of energy deposited in the electromagnetic calorimeter. Candidates without a matching track or 
reconstructed conversion vertex in the inner detector are classified as unconverted photons~\cite{PERF-2013-04}. Those with a 
matching reconstructed conversion vertex or a matching track consistent with originating from a photon conversion are classified as converted photons. Photons 
reconstructed within $|\eta^{\gamma}|<$ 2.37 are retained. Those in the transition region between the barrel and end-caps (1.37 $<|\eta^{\gamma}|<$ 1.56) or regions of the calorimeter affected by read-out or high-voltage failures are not considered in the event reconstruction.

Photon candidates passing loose identification requirements, based on the energy leaking into the hadronic calorimeter and 
the lateral shower shape in the second layer of the electromagnetic calorimeter, are retained~\cite{STDM-2011-04,STDM-2012-16}. The photon cluster energies are corrected using an in situ calibration based on the $Z\rightarrow e^{+}e^{-}$ reconstructed mass peak~\cite{PERF-2013-05}. Once these corrections are applied, the three reconstructed photons with the highest transverse energies $E_{\text T}^{\gamma_{1}}$, $E_{\text T}^{\gamma_{2}}$ and $E_{\text T}^{\gamma_{3}}$ in each event are retained. Events with $E_{\text T}^{\gamma_{1}}$, $E_{\text T}^{\gamma_{2}}$ and $E_{\text T}^{\gamma_{3}}$ greater than 27~GeV, 22~GeV and 15~GeV, respectively, and with a $\Delta R$ distance in the $\eta$--$\phi$ plane above 0.45 between pairs of photons, are selected. Additionally, the invariant mass of the triphoton system $m^{\gamma\gamma\gamma}$ is required to be above 50~GeV. This requirement corresponds to the minimum value of $m^{\gamma\gamma\gamma}$ predicted at particle level by 
the signal MC sample described in Section~\ref{sec:mc_simulations}.

Two further criteria are used to define the signal region and the background-enriched regions used to estimate the jet-to-photon misidentification background. A tight photon-identification selection~\cite{PERF-2013-04} is applied to reject hadronic jet background, by imposing requirements on nine discriminating variables (referred to as ``shower shapes'') computed from the energy leaking into the hadronic calorimeter and the lateral and longitudinal shower development in the electromagnetic calorimeter. The efficiency of this selection for one photon is $\approx$ 67\% ($>$ 90\%) for $E_{\text T}^{\gamma}$ $\approx$ 15 GeV ($>$ 100 GeV). For the MC simulations, the shower-shape 
variables are shifted to correct for small differences in the average values between data and the simulation. In addition, $E_{\text T}^{\gamma}$- and 
$\eta^{\gamma}$-dependent factors are applied to correct for the residual mismatch between the photon identification efficiencies in the simulation 
and the data. The isolation of the photon $E_{\text T}^{\text{iso}}$ is based on the amount of transverse energy inside a cone of size $\Delta R$ = 0.4 in the $\eta$-$\phi$ plane 
around the photon candidates, excluding an area of size $\Delta\eta\times\Delta\phi$ = 0.125 $\times$ 0.175 centred on the photon energy cluster. The isolation 
transverse energy is computed from the topological clusters of calorimeter cells~\cite{PERF-2014-07}. The measured $E_{\text T}^{\text{iso}}$ is corrected 
for the leakage of the photon's energy into the isolation cone and the estimated contributions from the UE and pile-up. These latter two corrections are computed simultaneously on an 
event-by-event basis and the combined correction is typically between 1.5 and 2.0 GeV~\cite{STDM-2014-09}. The $E_{\text T}^{\text{iso}}$ value for isolated photons is required to be lower than $E_{\text T}^{\text{iso}}$~=~0.025$\cdot E_{\text T}^{\gamma}$~+~2.7~[GeV]. The efficiency of the isolation requirement is typically above 80\% and increases as a function of $E_{\text T}^{\gamma}$.~The number of data events selected in the signal region is 1085. For background studies, two alternative categories of photons are defined. First, non-tight photon candidates are defined as those passing the loose selection but not satisfying the tight identification criteria for at least one of the shower-shape variables computed from the energy deposits in cells of the first layer of the EM calorimeter. Second, non-isolated photon candidates are defined to have $E_{\text T}^{\text{iso}}>$~0.025$\cdot~E_{\text T}^{\gamma}$~+~4.7~[GeV].
\section{Background estimation and signal extraction}
\label{sec:bkg_estimation}

The background contributions to the signal come from high-$p_{\text T}$ jets and electrons that are misidentified as isolated photons (referred to as jet and electron backgrounds). The estimation of these backgrounds is explained in the following.

\subsection{$e$--$\gamma$ misidentification}
\label{sec:bkg_estimation_e}

The number of background events due to $e$--$\gamma$ misidentification is estimated using the MC samples listed in Section~\ref{sec:mc_description}. 
The {\textsc Sherpa} MC events were weighted to correct the $e$--$\gamma$ misidentification rates to match those found in data 
(referred to as $e$--$\gamma$ scale factors in the following). These weights were estimated from $Z\rightarrow e^{+}e^{-}$ events where at least either the 
electron or the positron was reconstructed as a photon. The expected number of electron background events in the signal region is 71$\pm$2 (stat), which corresponds to (6.5$\pm$0.2)\% of the selected events. A systematic uncertainty is computed by propagating the uncertainty in the 
$e$--$\gamma$ scale factors to the estimation of the yield (see Section~\ref{sec:sys_uncert}).

The normalisation of the MC samples is tested by fitting the signal, $e$--$\gamma$ and jet--$\gamma$ misidentification 
contributions to the data as a function of $m^{\gamma\gamma\gamma}$ in the region 50~$< m^{\gamma\gamma\gamma}<$~125~GeV. Since 86\% of 
electron background events come from processes where a photon is emitted by an electron or positron originating from the decay of a $Z$ boson 
($pp\rightarrow Z\rightarrow e^{+}e^{-}\gamma$), a peak around $m^{\gamma\gamma\gamma}\approx$ $m_{Z}$ is expected. To enhance the relative contribution of electrons that are misidentified as photons, only events with at least one converted photon are considered. Signal and 
electron background MC events are used to describe the shape of the $m^{\gamma\gamma\gamma}$ distribution, whereas data events with at least one non-tight identified photon are used to describe the jet background contribution. The fit gives an electron background yield that is 
consistent with the MC estimation, since it predicts a correction factor equal to 1.0 $\pm$ 0.4 (stat). Moreover, the result of the fit is found to be independent of the definition of non-tight identified photons and a change of $<$ 2\% is found 
when the isolation requirement is loosened by 1 GeV. 

\subsection{Jet--$\gamma$ misidentification}
\label{sec:bkg_estimation_jet}

A large background from jet--$\gamma$ misidentification remains in the selected sample, even after imposing the tight identification 
and isolation requirements on the photons. The jet background originates from multi-jet ($jjj$), photon + jets ($\gamma jj$), and 
diphoton + jets ($\gamma\gamma j$) processes in which at least one jet is misidentified as a photon. The two-dimensional-sideband method exploited in 
Refs.~\cite{STDM-2010-08,STDM-2011-05,STDM-2012-05,STDM-2012-16,STDM-2012-18,STDM-2014-09} to measure the inclusive photon and diphoton differential cross sections is used to perform an in situ statistical subtraction of the background.~The method uses the photon 
isolation energy and photon identification criteria to discriminate prompt photons from jets. It relies on the fact that the correlations between the isolation and identification variables in jet background events are small, and that the signal contamination in the non-tight 
or non-isolated control region is low.

The two-dimensional-sideband method counts all combinations of photons meeting or failing to meet the tight identification or isolation criteria. Four categories are defined for each photon, resulting in 64 categories of events where 63 of these categories correspond to $jjj$, $\gamma jj$, and $\gamma\gamma j$ background-enriched regions.~The inputs of the method are the number of events in each category, the correlation between the isolation and identification variables in jet background events ($R^{\text{bg}}$), the signal leakage fractions in non-tight and non-isolated regions, and the expected number of electron background events in each category.~The correlation between the isolation and identification variables is taken to be negligible ($R^{\text{bg}}$ = 1.0) based on studies in simulated background samples and on data in a background-dominated region~\cite{STDM-2014-09}. The signal leakage fractions and electron-background events are estimated using the MC samples described in Section~\ref{sec:mc_description}.

The method allows the extraction of the number of true three-photon signal events ($N^{\gamma\gamma\gamma}$), the number of events where 
at least one, two and three candidates are true jets and the tight and isolation efficiencies for fake 
photon candidates from jets (``fake rates''). The number of events in each category is expressed as a function of the following parameters: signal, electron- and jet-background yields, signal leakage fractions, fake rates, and $R^{\text{bg}}$. Then, the system of 64 independent equations is grouped into 21 dependent linear equations which are solved iteratively using a $\chi^{2}$ minimisation procedure. The size of each bin of the observables under study is chosen to have a sufficiently large number of events to apply this method bin-by-bin. The statistical uncertainty of the signal and jet background-enriched regions is propagated to the estimation of the three-photon signal yield via pseudo-experiments.

The signal purity, defined as $N^{\gamma\gamma\gamma}/N_{\text{SR}}$, where $N_{\text{SR}}$ is the number of selected events in the signal region, is found to be (55$\pm$5)\%~(stat), with a value of $\approx$ 45\% ($\approx$ 60\%) at low (high) $E_{\text T}^{\gamma}$. The fractions of $\gamma\gamma j$, $\gamma jj$ and $jjj$ events are (33$\pm$2)\%~(stat), (5$\pm$2)\%~(stat) and (0.2$\pm$0.2)\%~(stat) respectively. Systematic uncertainties are assigned to the modelling of the non-tight and non-isolated signal leakage fractions and to the value of $R^{\text{bg}}$ (see Section~\ref{sec:sys_uncert}). 
\section{Unfolding to particle level}
\label{sec:unfolding}

The production cross section for three isolated photons is measured as functions of $E_{\text T}^{\gamma_{1}}$, $E_{\text T}^{\gamma_{2}}$, $E_{\text T}^{\gamma_{3}}$, $\Delta\phi^{\gamma_{1}\gamma_{2}}$, $\Delta\phi^{\gamma_{1}\gamma_{3}}$, $\Delta\phi^{\gamma_{2}\gamma_{3}}$, $|\Delta\eta^{\gamma_{1}\gamma_{2}}|$, $|\Delta\eta^{\gamma_{1}\gamma_{3}}|$, $|\Delta\eta^{\gamma_{2}\gamma_{3}}|$, $m^{\gamma_{1}\gamma_{2}}$, $m^{\gamma_{1}\gamma_{3}}$, $m^{\gamma_{2}\gamma_{3}}$ and $m^{\gamma\gamma\gamma}$. The fiducial phase-space region is listed in Table~\ref{unf:tab1}. The predictions of the MC generators at particle level are defined using those particles with a lifetime $\tau$ longer than 30 ps; 
these particles are referred to as ``stable''. The particles associated with the overlaid $pp$ collisions are not considered. The particle-level isolation requirement on the photons is built by summing the transverse energy of all stable particles, except for muons and neutrinos, in a cone of size $\Delta R$ = 0.4 around the photon direction. The contribution from the UE is subtracted using the same procedure as applied to the data at the reconstruction level~\cite{jhep:1608.005}.~The data distributions after background subtraction are unfolded to the particle level using bin-by-bin correction factors determined using the signal MC sample.~The correction factors take into account the efficiency of the event and photon selection criteria and the small migration effects. Of the signal events reconstructed in a given bin, the fraction that are generated in the same bin is typically found to be $>$ 93\%. The data distributions are unfolded to the particle level via the formula

 \begin{equation}\nonumber
   \frac{{\text d}\sigma}{{\text d}A} (i) = \frac{N^{\text{sig}}(i)C(i)}{\Delta A(i)~{\cal L}},
  \end{equation}
  
 where for a given bin $i$, $({\text d}\sigma/{\text d}A)$ is the differential cross section as a function of observable $A$, $N^{\text{sig}}$ is the number of 
 background-subtracted data events, $C$ is the correction factor, ${\cal L}$ is the integrated luminosity and
 $\Delta A$ is the width of the bin. The correction factors are computed using the MC sample of events as  
 $C(i) = N_{\text{part}}^{\text{MC}}(i)/N_{\text{reco}}^{\text{MC}}(i)$, where $N_{\text{part}}^{\text{MC}}(i)$ is the number of events which satisfy the kinematic 
 constraints of the phase-space region at the particle level, and $N_{\text{reco}}^{\text{MC}}(i)$ is the number of events which fulfil all the selection criteria 
 at the reconstruction level. The correction factors vary between 1.5 and 3.3 as functions of photon transverse energy, invariant mass of pairs of photons, 
 and the invariant mass of the triphoton system, whereas they have a constant value close to 2.5 as functions of the difference in azimuthal angle and 
 in pseudorapidity between pairs of photons.
 
\begin{table}[t]
\caption{Fiducial phase-space region defined at particle level.}
\begin{center}
\begin{tabular}{c | c }
\hline
\hline
{{\small Requirements on}} & {\small{Phase-space region}} \\
\hline
{\small $E_{\text T}^{\gamma}$}  & {\small $E_{\text T}^{\gamma_{1}}>$ 27 GeV, $E_{\text T}^{\gamma_{2}}>$ 22 GeV, $E_{\text T}^{\gamma_{3}}>$ 15 GeV}\\
\hline
{\small $m^{\gamma\gamma\gamma}$}  &{\small  $m^{\gamma\gamma\gamma}>$ 50 GeV}\\
\hline
{\small $\Delta R^{\gamma\gamma}$}  &{\small  $\Delta R^{\gamma\gamma}>$ 0.45}\\
\hline
{\small $|\eta^{\gamma}|$}  &{\small  $|\eta^{\gamma}|<$ 2.37 (excluding 1.37 $<|\eta^{\gamma}|<$ 1.56)}\\ 
\hline
{\small Isolation} &{\small  $E_{\text T}^{\text{iso}}<$ 10 GeV} \\
\hline
\hline
\end{tabular}
\label{unf:tab1}
\end{center}
\end{table}
\section{Experimental and theoretical uncertainties}
\label{sec:sys_uncert}

\subsection{Experimental uncertainties}
\label{sec:sys_uncert_exp}

The sources of experimental systematic uncertainty that affect the measurements are the photon energy scale and resolution, photon identification, 
jet and electron background subtraction, modelling of the photon isolation, the photon fragmentation contribution, the unfolding procedure and the luminosity.
\begin{itemize}
 \item \textbf{Photon energy scale and resolution.} The uncertainty due to the photon energy scale is estimated by varying the photon energies in the MC simulation~\cite{PERF-2013-05}. This uncertainty mostly affects the $C(i)$ correction factor. The effect of this variation on the 
 estimation of the cross section is typically $<$~2\%. In addition, the uncertainty in the energy resolution is estimated by smearing photon energies in the MC simulation as described in Ref.~\cite{PERF-2013-05}. The resulting uncertainty in the cross section is typically $<$ 0.1\%.
 \item \textbf{Photon identification efficiency.} The uncertainty in the photon identification efficiency is estimated from the effect of 
differences between shower-shape variable distributions in data and simulation~\cite{PERF-2013-04}. This uncertainty affects the estimation of the non-tight signal leakage fractions and the $C(i)$ correction factor and is fully correlated between photons. The correlation between tight and non-tight identification variables is also considered in the propagation of the uncertainty. The resulting uncertainty in the cross section is $\approx$ 10\% ($\approx$ 4\%) at low (high) $E_{\text T}^{\gamma}$.
 \item \textbf{Photon identification and isolation correlation in the background.} The photon isolation and identification variables used to define the 
  two-dimensional background sidebands are assumed to be independent in jet background events ($R^{\text{bg}}$ = 1.0). Any correlation 
  between these variables affects the estimation of signal purity and leads to systematic uncertainties in the background-subtraction procedure. The value of $R^{\text{bg}}$ is estimated using background MC samples and is found to be consistent with unity within $\pm$10\%~\cite{STDM-2012-18,STDM-2014-09}. This value of $R^{\text{bg}}$ is verified using background-enriched regions in data. The assumption of $R^{\text{bg}}$ = 1.0 is found to hold within $\pm$10\% in the kinematic region of the measurements presented here. The resulting uncertainty in the cross section is $\approx$ 8\% ($\approx$ 4\%) at low (high) $E_{\text T}^{\gamma}$.
 \item \textbf{Photon isolation modelling.} Differences between data and signal MC events in the modelling of the isolation distribution can lead to systematic uncertainties in the estimation of the non-isolated signal leakage fractions and the $C(i)$ correction factor. Two subsamples are selected from data by applying either the tight or non-tight identification criteria to each photon; the subsample selected with non-tight identification criteria is expected to be enriched in background candidates. The $E_{\text T}^{\text{iso}}$ value for the non-tight candidates is scaled so that the integral for $E_{\text T}^{\text{iso}}>$ 10 GeV, where the contribution from the signal is expected to be negligible, matches that of the tight candidates. The rescaled background distribution is subtracted from that of the tight photon candidates to extract the isolation profile of signal-like candidates. These distributions are used to derive Smirnov transformations~\cite{PERF-2013-04}. The Smirnov transformation shifts the photon isolation values event-by-event in MC simulation to match the isolation distribution found in data. This Smirnov-transformed MC sample is used to estimate new differential cross sections. Differences from the nominal results are taken as systematic uncertainties. The resulting uncertainty in the cross section is $\approx$~7\% ($\approx$ 4\%) at low (high) $E_{\text T}^{\gamma}$.
 \item \textbf{Photon fragmentation contribution.} The admixture of direct and fragmentation photons affects the estimation of the signal leakage fractions which are used in the jet background subtraction procedure and the $C(i)$ correction factor. A photon originating from the fragmentation of a parton can be modelled in the MC simulation by allowing the radiation of a photon by a parton. A sample of fragmentation photons is selected by applying the event selection to a diphoton MC sample (see Section~\ref{sec:mc_description}). This selects three-photon events where at least one of the final-state photons results from fragmentation. The diphoton MC sample predicts that for more than 98\% of the events the sub-sub-leading photon originates from parton bremsstrahlung. Differences in the isolation distributions between direct and fragmentation photons are expected. Therefore, a template fit to the sub-sub-leading photon isolation distribution is performed to determine the optimal admixture of the nominal and diphoton MC samples. The direct and fragmentation isolation templates are given by the nominal and diphoton MC samples respectively, whereas the jet background template is taken from a data control region where the sub-sub-leading photon candidate satisfies the non-tight selection. The fit estimates that about 40\% of the sub-sub-leading photons originate from fragmentation, as modelled by the diphoton MC sample. This value is used to merge the nominal and diphoton MC samples. The new MC sample is used to estimate the signal leakage fractions and the $C(i)$ correction factors. The deviation of the differential cross section from the value obtained using the Smirnov-transformed MC sample is taken as the systematic uncertainty. This avoids double counting the effect of the photon isolation modelling. The resulting uncertainty in the cross section is $\approx$ 4\%. 
 \item \textbf{ $e$--$\gamma$ misidentification.} The uncertainty in the electron background contamination is estimated by propagating the uncertainty in the $e$--$\gamma$ scale factors (see Section~\ref{sec:bkg_estimation_e}), which affects the prediction of the $e$--$\gamma$ misidentification rates, to the estimation of the cross section. The resulting uncertainty is $\approx$~0.1\%.
 \item \textbf{Unfolding procedure.} The effect of unfolding is investigated by using smooth functions to re-weight the signal MC simulation to match the data distributions after background subtraction. The data are unfolded using this reweighted MC sample and the resulting cross sections are compared to the nominal measurements. The differential cross sections are found to differ by < 1\%.
 \item \textbf{Other sources.} The effect of different amounts of pile-up is estimated by comparing the ratio of data to MC simulated signal for high and low pile-up samples. No dependence of this ratio on pile-up conditions is found. In addition, the effect of the trigger efficiency on the estimation of the cross section is found to be $<$ 0.3\%. The uncertainty in the integrated luminosity is 1.9\%~\cite{DAPR-2013-01}.
\end{itemize}

 The total systematic uncertainty is computed by adding in quadrature the uncertainties from the sources listed above and is found to be $\approx$ 13\%. It 
 decreases as a function of $E_{\text T}^{\gamma}$ from $\approx$ 15\% to $\approx$ 10\%. For regions with $E_{\text T}^{\gamma_{1}}\gtrsim$ 50~GeV, $E_{\text T}^{\gamma_{2}}\gtrsim$ 50~GeV and $E_{\text T}^{\gamma_{3}}\gtrsim$ 30~GeV, the uncertainty of the measurements is dominated by the statistical uncertainty of the data. Table~\ref{sys:tab2} shows the breakdown of the systematic uncertainties in the 
 measurement of the inclusive fiducial cross section. The statistical uncertainty in the measured inclusive fiducial cross section is $\approx$ 9\%.

\begin{table}[t]
\caption{Breakdown of the relative systematic uncertainties in the measurement of the inclusive fiducial cross section.}
\begin{center}
\begin{tabular}{c | c }
\hline
\hline
{\small \textbf{Source}} & {\small\textbf{Relative systematic uncertainty}}\\
\hline
{\small Photon identification efficiency} & 7.9\% \\ 
{\small Identification and isolation correlation in the background} & 7.7\% \\ 
{\small Photon isolation modelling} & 5.8\% \\
{\small Photon fragmentation contribution} & 3.9\% \\
{\small Photon energy scale and resolution} & 1.6\% \\
{\small Unfolding} & 0.6\% \\
{\small $e$--$\gamma$ misidentification} & 0.1\% \\
{\small Measurement of the integrated luminosity} & 1.9\% \\
\hline
{\small Total} & 13\% \\
\hline
\hline
\end{tabular}
\label{sys:tab2}
\end{center}
\end{table}

\subsection{Theoretical uncertainties}
\label{sec:sys_uncert_the}

The following sources of uncertainty in the theoretical predictions are considered for the {\textsc MCFM} and {{\textsc MadGraph5}\_aMC@NLO} calculations.
\begin{itemize}
  \item The uncertainty in the NLO QCD calculations due to terms beyond NLO is estimated by repeating the calculations using values of 
  $\mu_{\text R}$, $\mu_{\text F}$ and $\mu_{\text f}$ scaled by factors 0.5 and 2.~For the {{\textsc MadGraph5}\_aMC@NLO} calculations, only the $\mu_{\text R}$ 
  and $\mu_{\text F}$ scales are varied. In addition, the scales are either varied simultaneously, individually or by fixing one and varying the other two. The final 
  uncertainty is taken as the largest deviation of the possible variations with respect to the nominal value.
  \item The uncertainty in the NLO QCD calculations due to uncertainties in the proton PDFs is estimated by repeating the calculations using the 52 
  additional sets from the CT10 error analysis~\cite{PhysRevD:82:2010}.
  \item The uncertainty in the NLO QCD calculations due to the value of $\alpha_{\text S}(m_{Z})$ = 0.118 is estimated by repeating the calculations 
 using two additional sets of proton PDFs~\cite{PhysRevD:82:2010} employing different values of $\alpha_{\text S}(m_{Z})$, namely 
 $\alpha_{\text S}(m_{Z})$ = 0.116 and 0.120.
\end{itemize}

The dominant theoretical uncertainty in the predicted cross section arises from the missing terms beyond NLO and amounts to 10--12\%. The uncertainty 
arising from the PDF variations amounts to 2--3\% and the uncertainty arising from the value of $\alpha_{\text S}(m_{Z})$ is below 2\%. The total theoretical uncertainty is obtained by adding in quadrature the individual uncertainties listed above and amounts to 10--13\%.
\section{Results}
\label{sec:result}

The measured inclusive fiducial cross section for the production of three isolated photons in the phase-space region given in Table~\ref{unf:tab1} is
\begin{center}
  $\sigma_{\text{meas}}$ = 72.6~$\pm$~6.5~({\text{stat.}})~$\pm$~9.2~({\text{syst.}})~fb,
\end{center}

where ``stat.'' and ``syst.'' denote the statistical and systematic uncertainties. The fiducial cross sections predicted at NLO by {\textsc MCFM} and
{{\textsc MadGraph5}\_aMC@NLO} are
\begin{center}
 $\sigma_{\text{NLO}}$ = 31.5~{}$^{+3.2}_{-2.5}$~{\text{fb}} ({\textsc MCFM}),
\end{center}

\begin{center}
 $\sigma_{\text{NLO+PS}}$  = 46.6~{}$^{+5.7}_{-3.6}$~{\text{fb}} ({{\textsc MadGraph5}\_aMC@NLO}).
\end{center}

The NLO QCD calculations underestimate the measured inclusive fiducial cross section by factors of 2.3 and 1.6 for {\textsc MCFM} and {{\textsc MadGraph5}\_aMC@NLO}, respectively.~The addition of the parton shower to the {{\textsc MadGraph5}\_aMC@NLO} prediction improves the agreement with the measured value. The NLO electroweak corrections are small and cannot account for the observed differences between NLO QCD and the measurements~\cite{EWtriphoton}. Similar discrepancies between the NLO calculations and the measurements are found for the prediction of the inclusive fiducial cross section for $\gamma\gamma$, $W\gamma\gamma$ and $Z\gamma\gamma$ production~\cite{STDM-2012-05,STDM-2013-05,STDM-2014-01}. The NNLO calculations, which are available for the computation of $\gamma\gamma$ but not for $\gamma\gamma\gamma$ production, significantly improve the description of the diphoton fiducial cross section~\cite{STDM-2015-15,Aaboud:2017vol,jhep07:2016:148}.

\begin{figure}[htp]
 \begin{center}
  \centering
  \hspace{0.0cm}\subfloat[]{\includegraphics[scale=0.40]{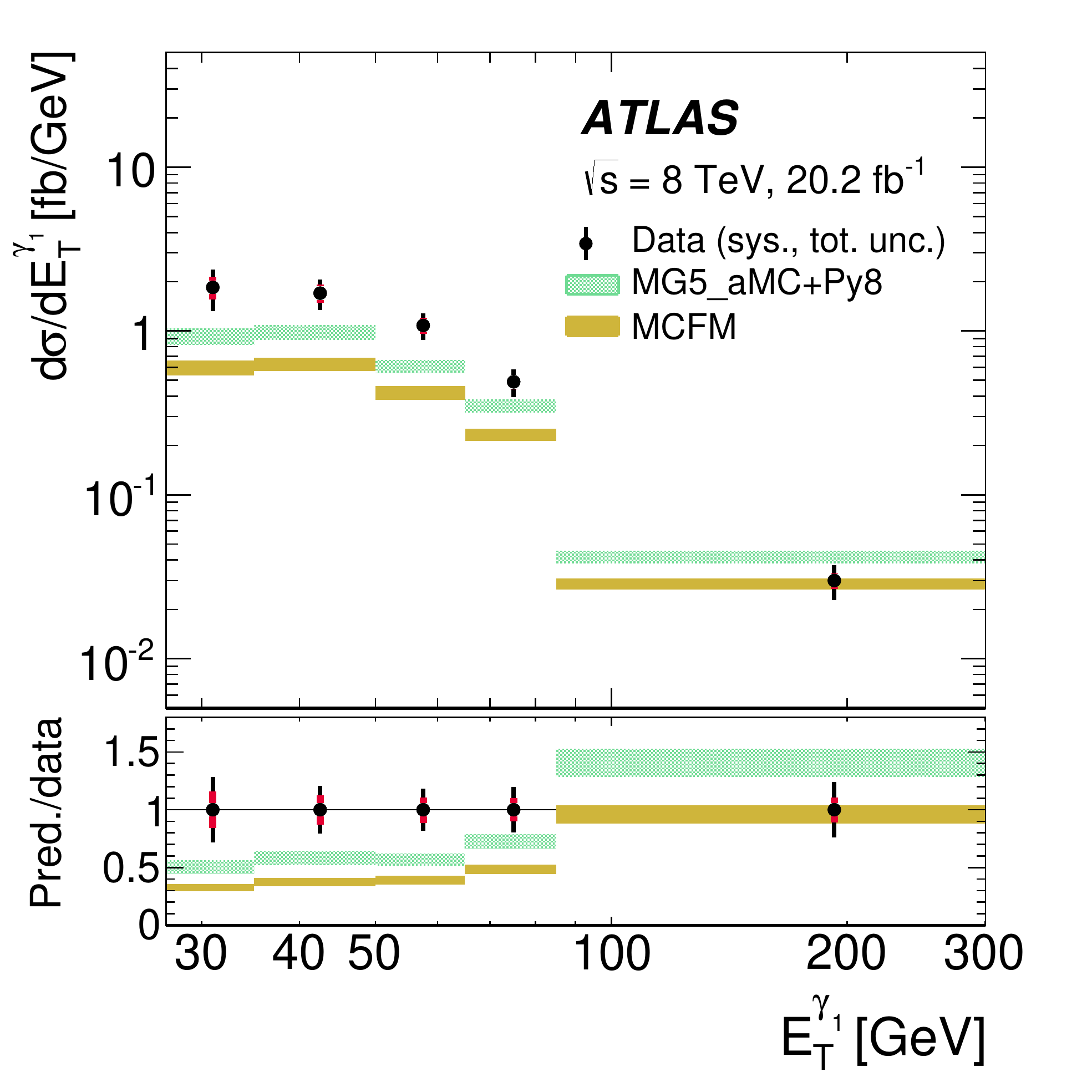}}
  \hspace{0.0cm}\subfloat[]{\includegraphics[scale=0.40]{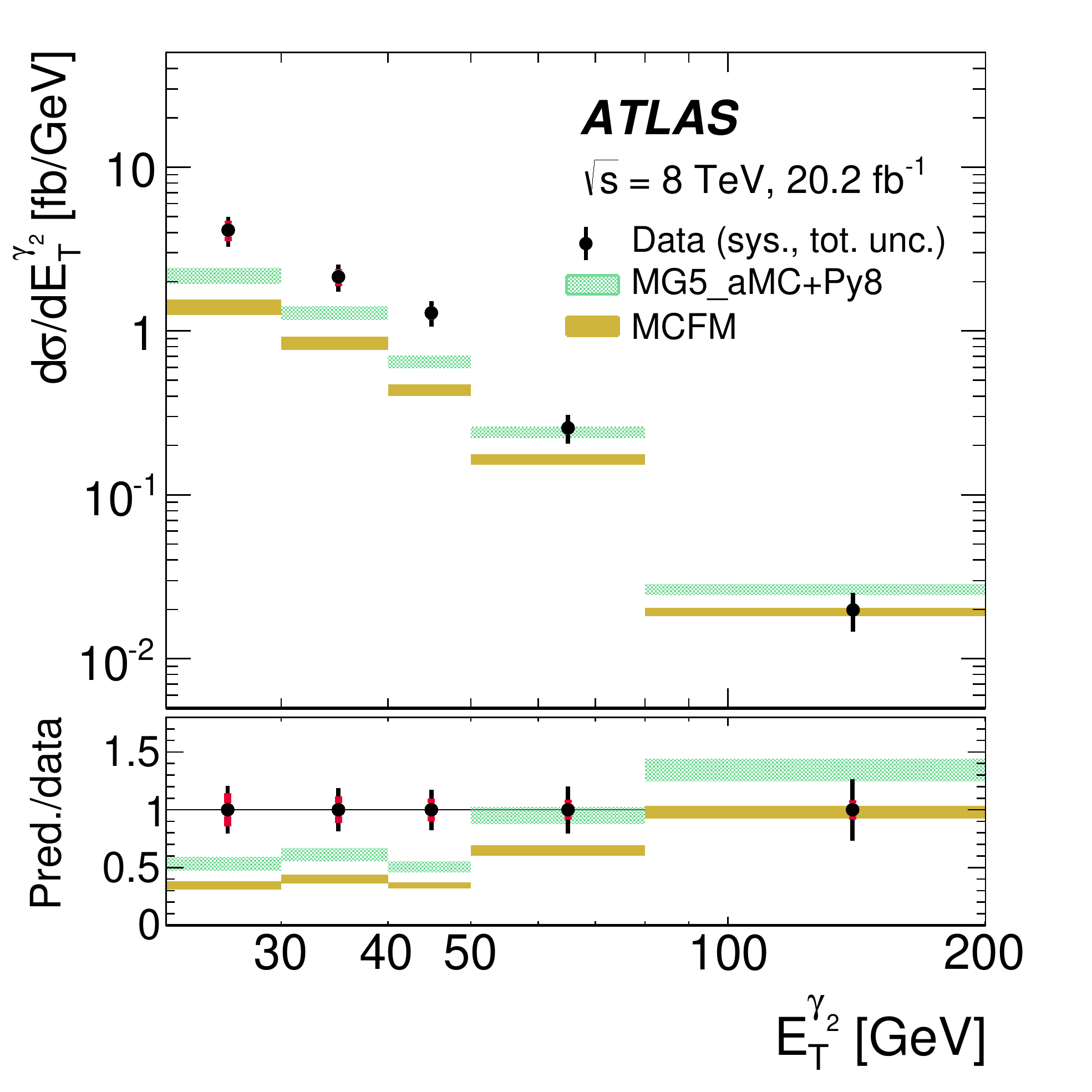}}
  \hspace{0.0cm}\subfloat[]{\includegraphics[scale=0.40]{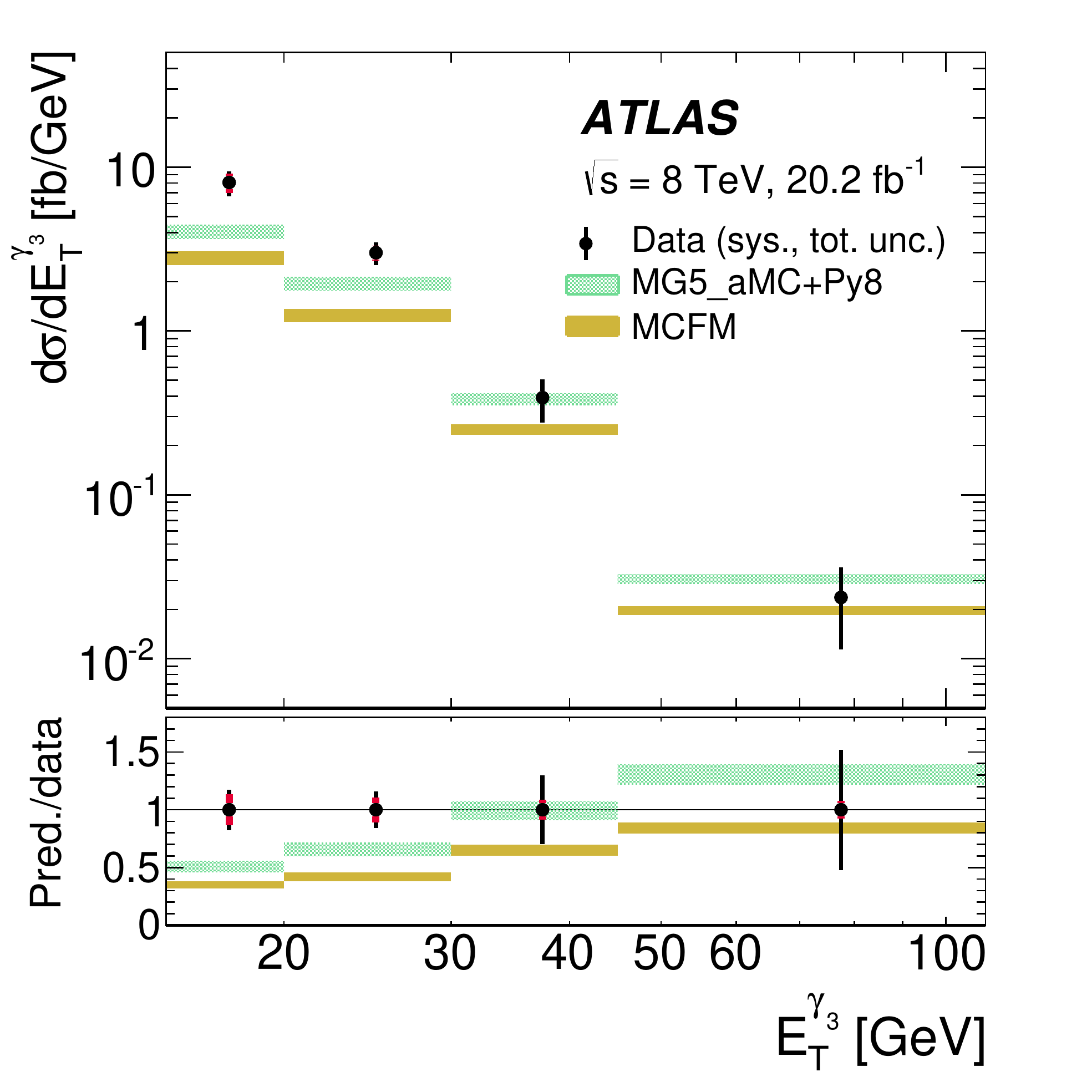}}
  \caption{Measured differential cross sections for the production of three isolated photons (dots) as functions of (a)~$E_{\text T}^{\gamma_{1}}$, (b)~$E_{\text T}^{\gamma_{2}}$ and (c)~$E_{\text T}^{\gamma_{3}}$. The NLO QCD calculations from {\textsc MCFM} and {{\textsc MadGraph5}\_aMC@NLO} are also shown. The thickness of each theoretical prediction corresponds to the theoretical uncertainty. The bottom part of each figure shows the ratios of predicted and measured differential cross sections. The red inner (black outer) error bars represent the systematic uncertainties (the statistical and systematic uncertainties added in quadrature). For most of the data points, the inner error bars are smaller than the marker size and thus not visible.}
  \label{fig:result_mcfm1}
\end{center}
\end{figure}

Figure~\ref{fig:result_mcfm1} shows the three-isolated-photons differential cross sections as functions of $E_{\text T}^{\gamma_{1}}$, 
$E_{\text T}^{\gamma_{2}}$ and $E_{\text T}^{\gamma_{3}}$. The measurements are compared to NLO QCD predictions from {\textsc MCFM} and {{\textsc MadGraph5}\_aMC@NLO}.~The NLO QCD calculations fail to describe the regions of low $E_{\text T}^{\gamma_{1}}$,
$E_{\text T}^{\gamma_{2}}$ and  $E_{\text T}^{\gamma_{3}}$. Differences of up to 60\% are observed between data and the predictions. The description of the measurements by the theory is improved at high $E_{\text T}^{\gamma}$. In particular, {{\textsc MadGraph5}\_aMC@NLO} calculations describe the measured cross sections for $E_{\text T}^{\gamma_{2}}\gtrsim $ 50 GeV and $E_{\text T}^{\gamma_{3}}\gtrsim $ 30~GeV within the statistical and systematic uncertainties, whereas MCFM describes the data only at the highest values of $E_{\text T}^{\gamma_{1}}$, $E_{\text T}^{\gamma_{2}}$ and $E_{\text T}^{\gamma_{3}}$.

\begin{figure}[htp]
 \begin{center}
  \centering
  \hspace{0.0cm}\subfloat[]{\includegraphics[scale=0.40]{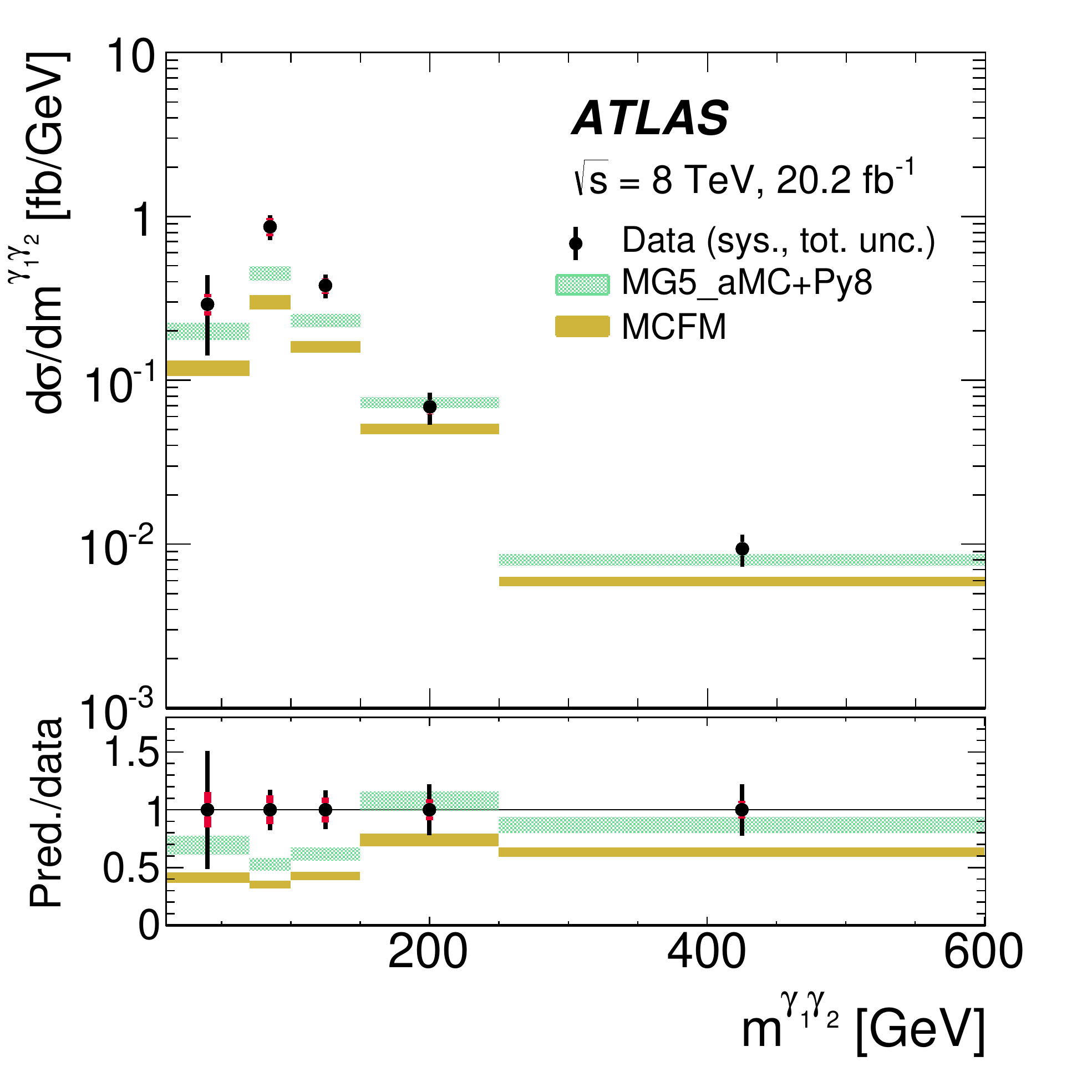}}
  \hspace{0.0cm}\subfloat[]{\includegraphics[scale=0.40]{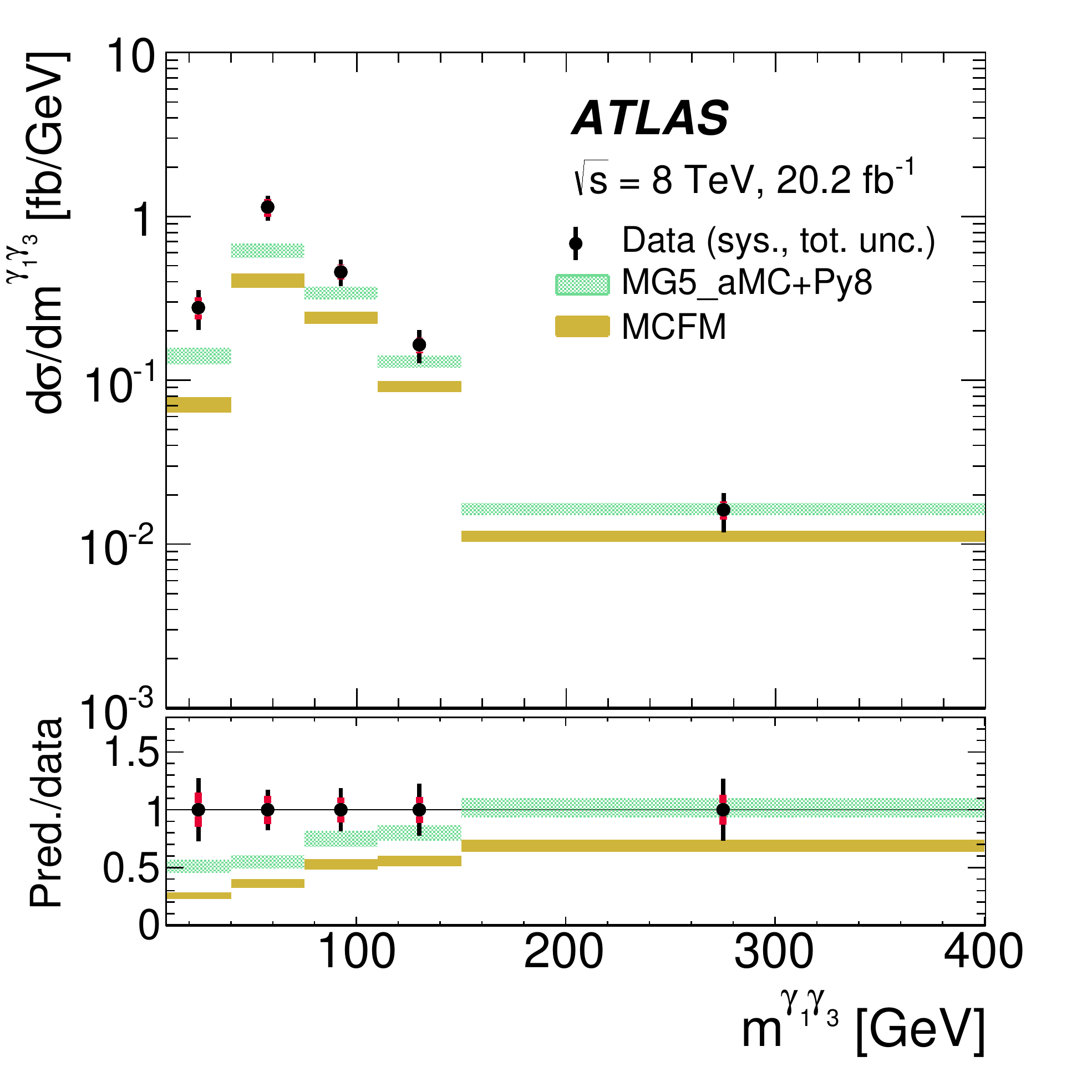}}
  \hspace{0.0cm}\subfloat[]{\includegraphics[scale=0.40]{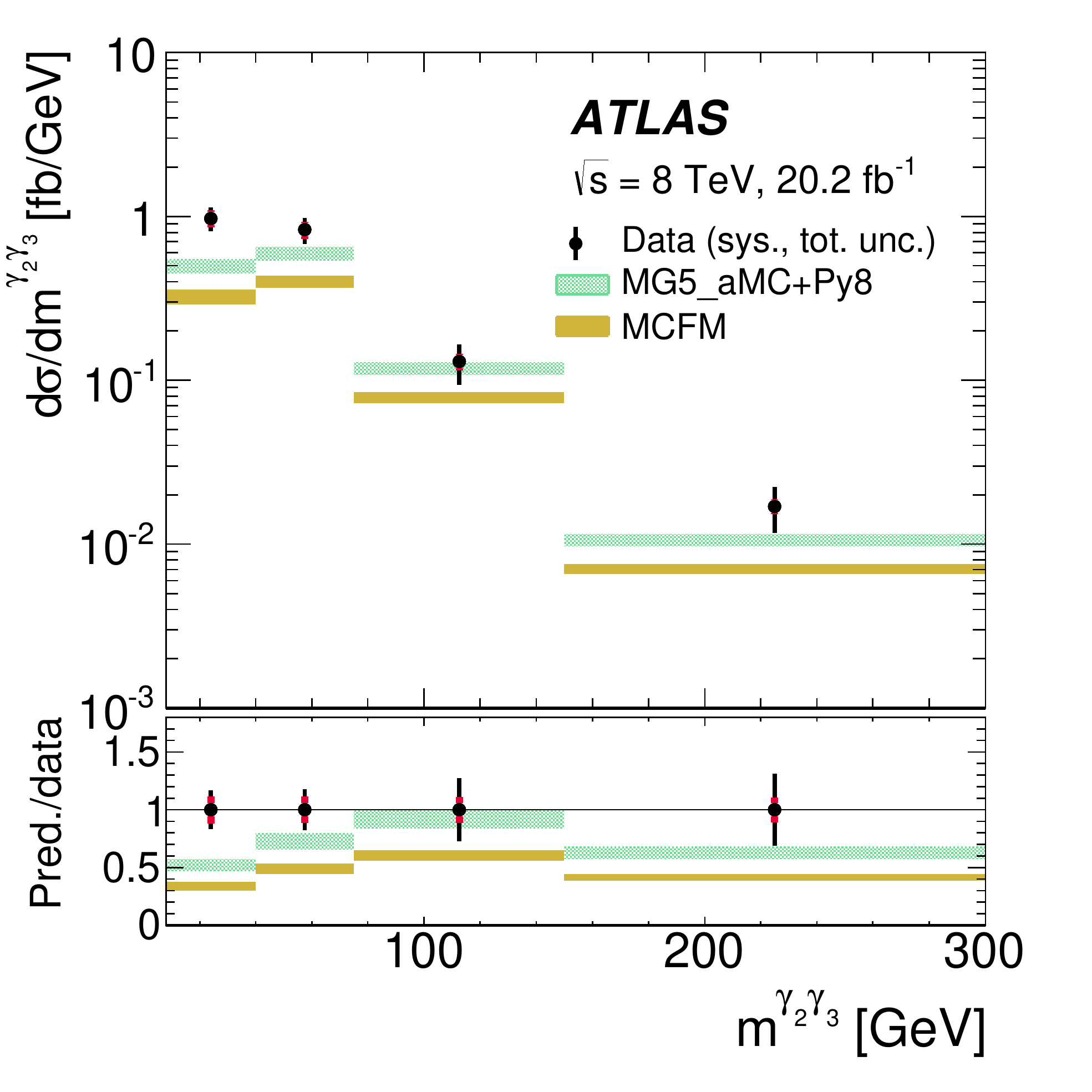}}
  \hspace{0.0cm}\subfloat[]{\includegraphics[scale=0.40]{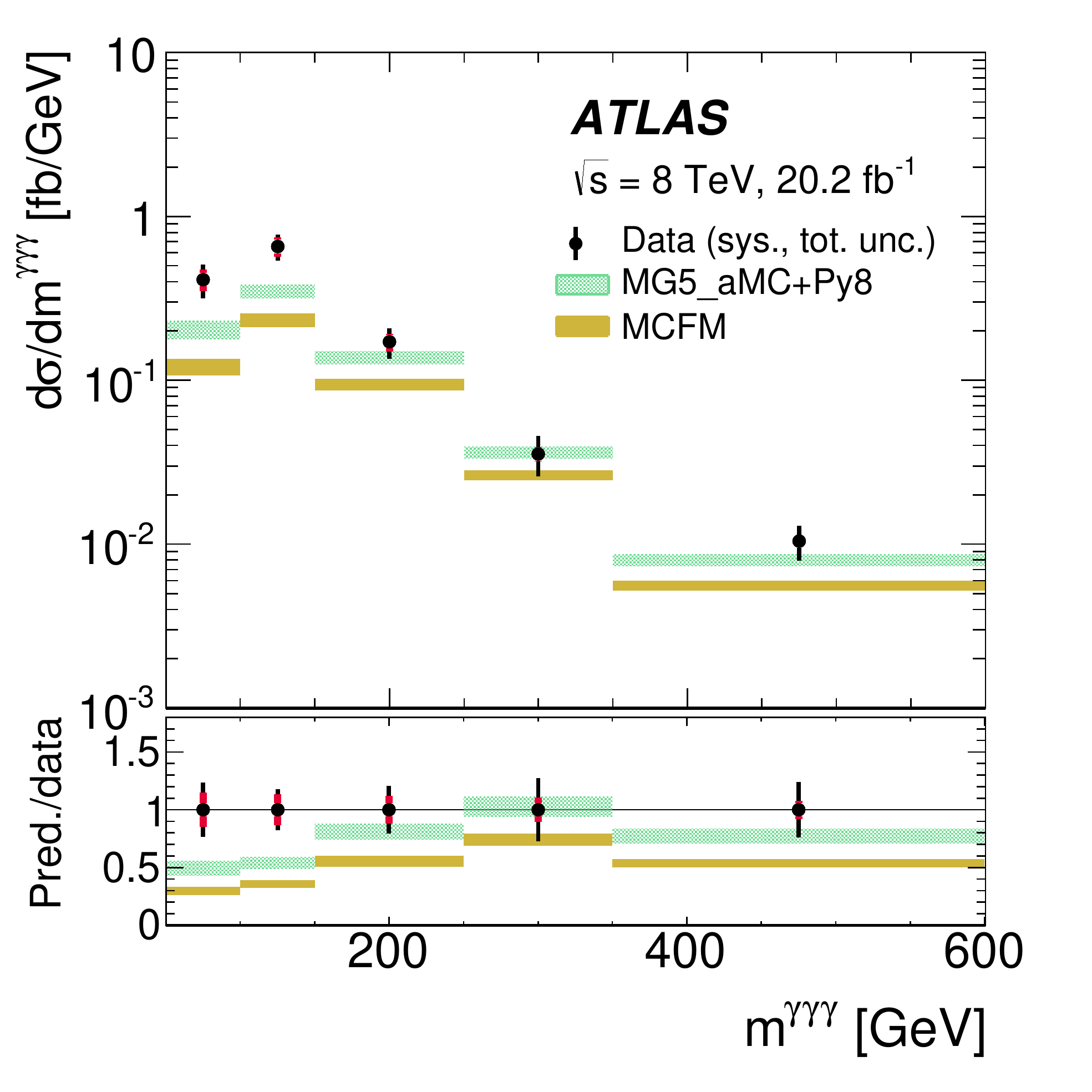}}
  \caption{Measured differential cross sections for the production of three isolated photons (dots) as functions of (a)~$m^{\gamma_{1}\gamma_{2}}$, (b)~$m^{\gamma_{1}\gamma_{3}}$, (c)~$m^{\gamma_{2}\gamma_{3}}$ and (d)~$m^{\gamma\gamma\gamma}$. The NLO QCD calculations from {\textsc MCFM} and {{\textsc MadGraph5}\_aMC@NLO} are also shown. The thickness of each theoretical prediction corresponds to the theoretical uncertainty. The bottom part of each figure shows the ratios of predicted and measured differential cross sections. The red inner (black outer) error bars represent the systematic uncertainties (the statistical and systematic uncertainties added in quadrature). For most of the data points, the inner error bars are smaller than the marker size and thus not visible.}
  \label{fig:result_mcfm2}
\end{center}
\end{figure}

A comparison of the NLO calculations to the measurements as functions of $m^{\gamma_{1}\gamma_{2}}$, $m^{\gamma_{1}\gamma_{3}}$, $m^{\gamma_{2}\gamma_{3}}$ and $m^{\gamma\gamma\gamma}$ is shown in Figure~\ref{fig:result_mcfm2}. The {\textsc MCFM} calculations underestimate the measurements by 50\% in the low invariant mass regions, whereas the differences are 30--40\% for $m^{\gamma_{1}\gamma_{2}} \gtrsim$~150~GeV, $m^{\gamma_{1}\gamma_{3}} \gtrsim$~75~GeV, $m^{\gamma_{2}\gamma_{3}} \gtrsim$~75~GeV and $m^{\gamma\gamma\gamma} \gtrsim$~150~GeV. The {{\textsc MadGraph5}\_aMC@NLO} calculations also underestimate the data by 30--50\% in the low invariant mass regions. However, they tend to give a better description of the measurements for $m^{\gamma_{1}\gamma_{2}} \gtrsim$~150~GeV, $m^{\gamma_{1}\gamma_{3}} \gtrsim$~75~GeV, $m^{\gamma_{2}\gamma_{3}} \gtrsim$~75~GeV and $m^{\gamma\gamma\gamma} \gtrsim$~150~GeV. For such regions, {{\textsc MadGraph5}\_aMC@NLO} predictions are 25--30\% higher than the {\textsc MCFM} estimates.

\begin{figure}[htp]
 \begin{center}
  \centering
  \hspace{0.0cm}\subfloat[]{\includegraphics[scale=0.305]{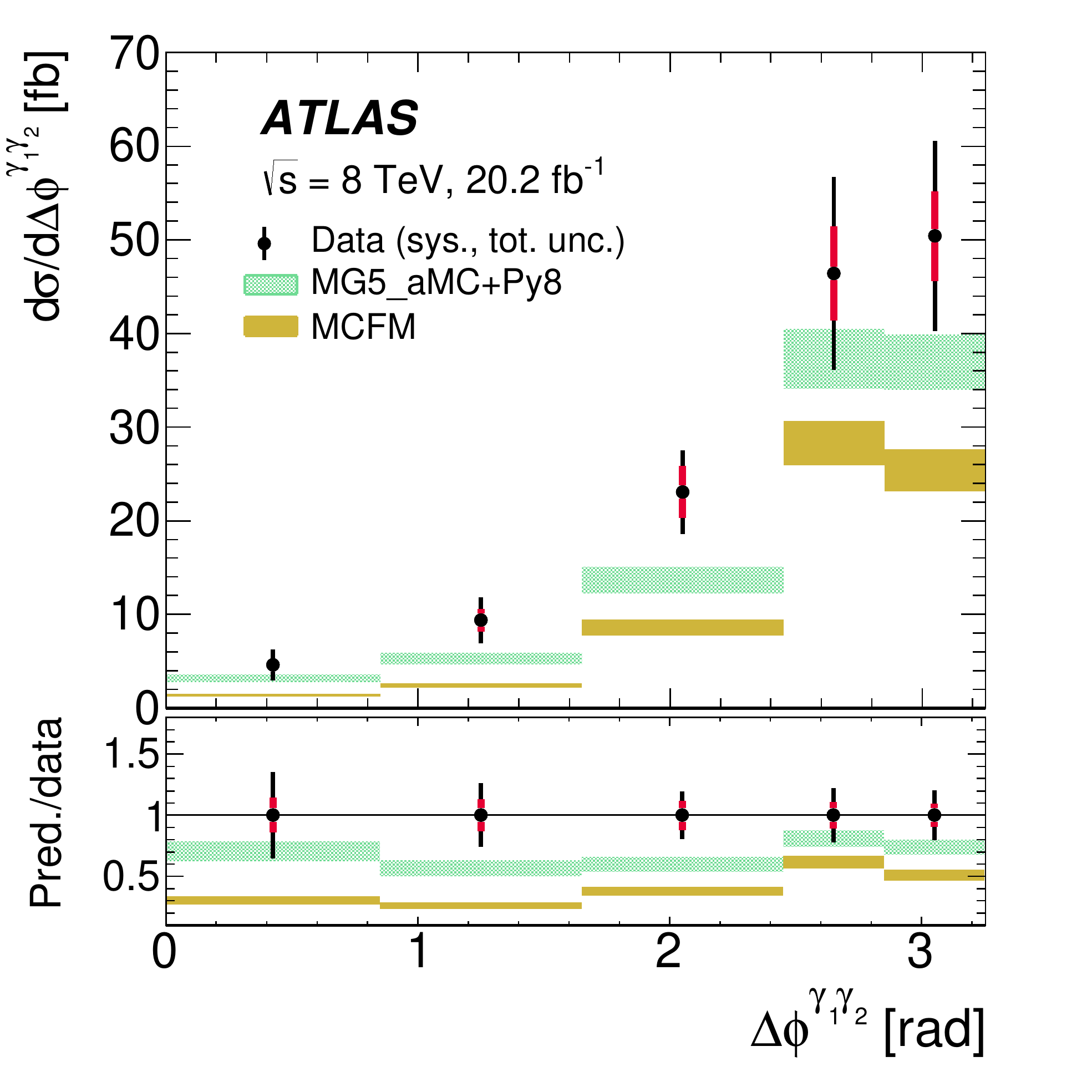}}
  \hspace{0.0cm}\subfloat[]{\includegraphics[scale=0.305]{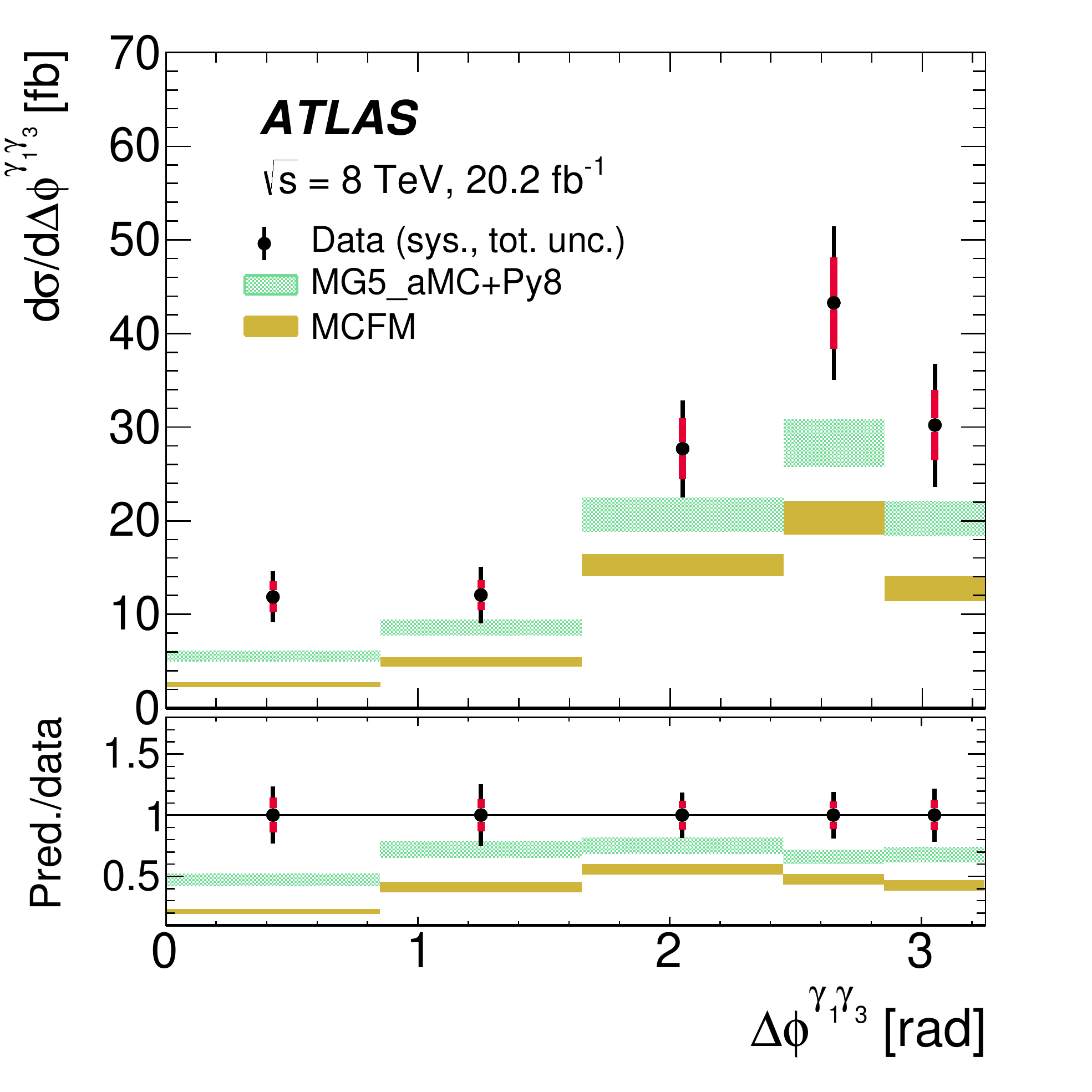}}\\
  \vspace{-0.3cm}
  \hspace{0.0cm}\subfloat[]{\includegraphics[scale=0.305]{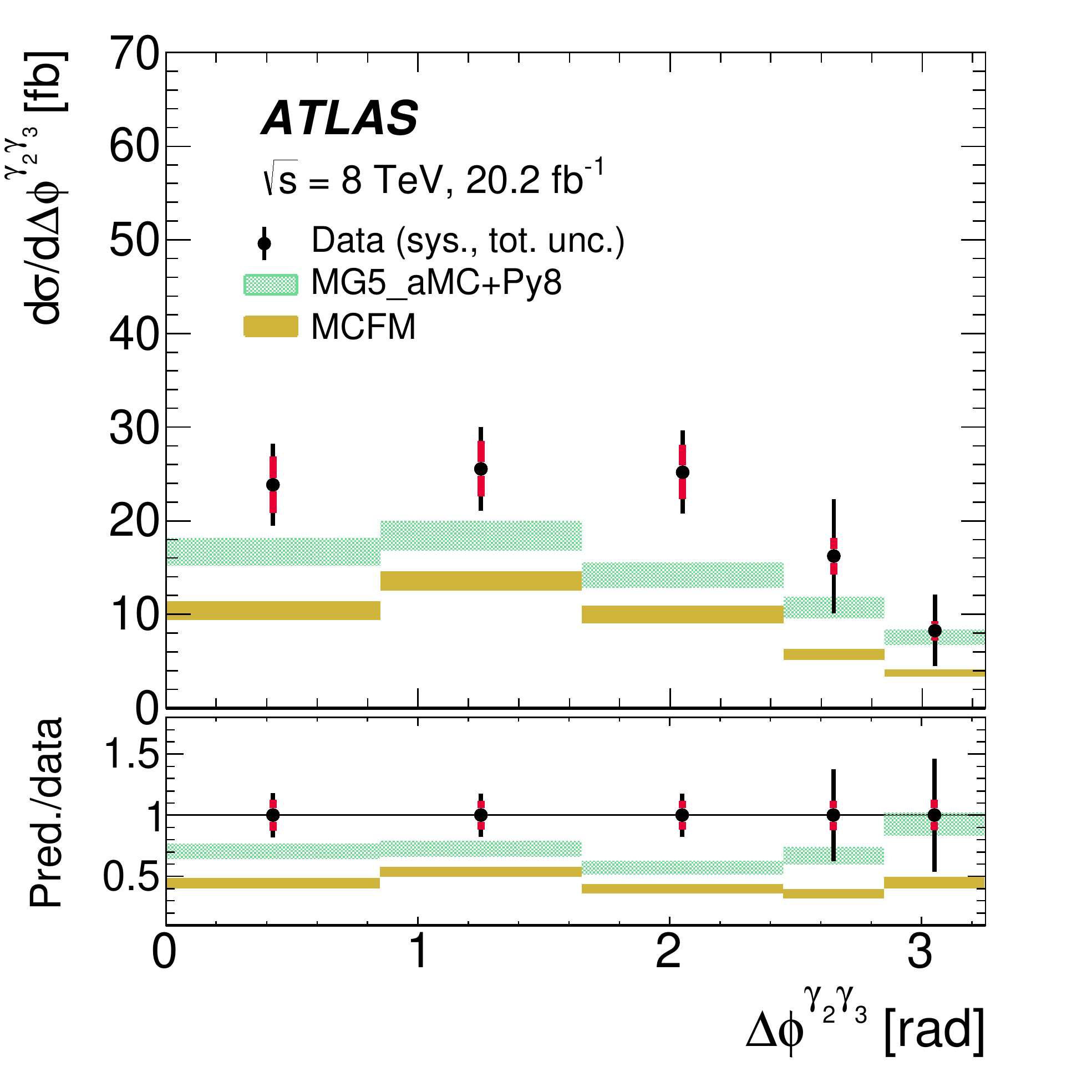}}
  \hspace{0.0cm}\subfloat[]{\includegraphics[scale=0.305]{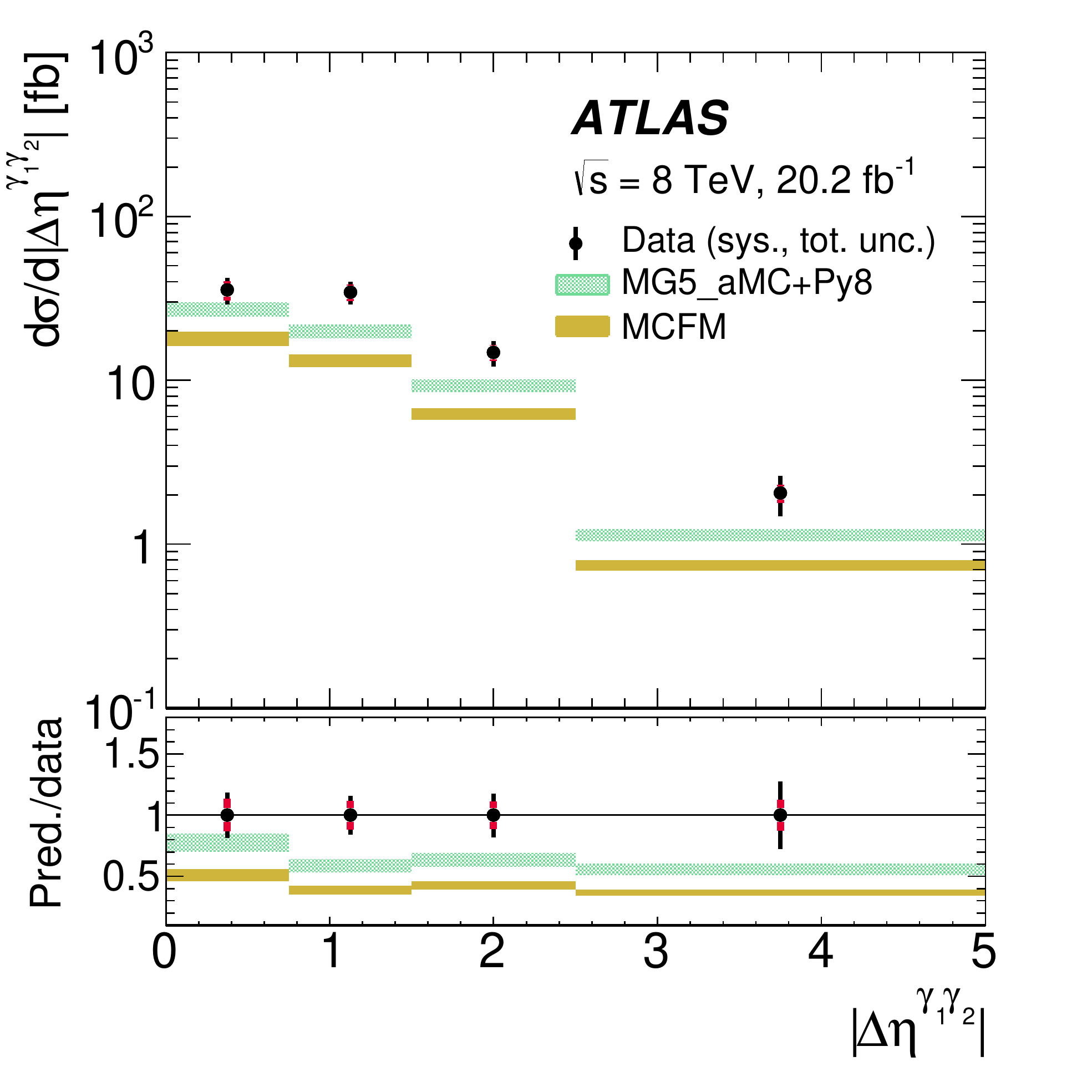}}\\
  \vspace{-0.3cm}
  \hspace{0.0cm}\subfloat[]{\includegraphics[scale=0.305]{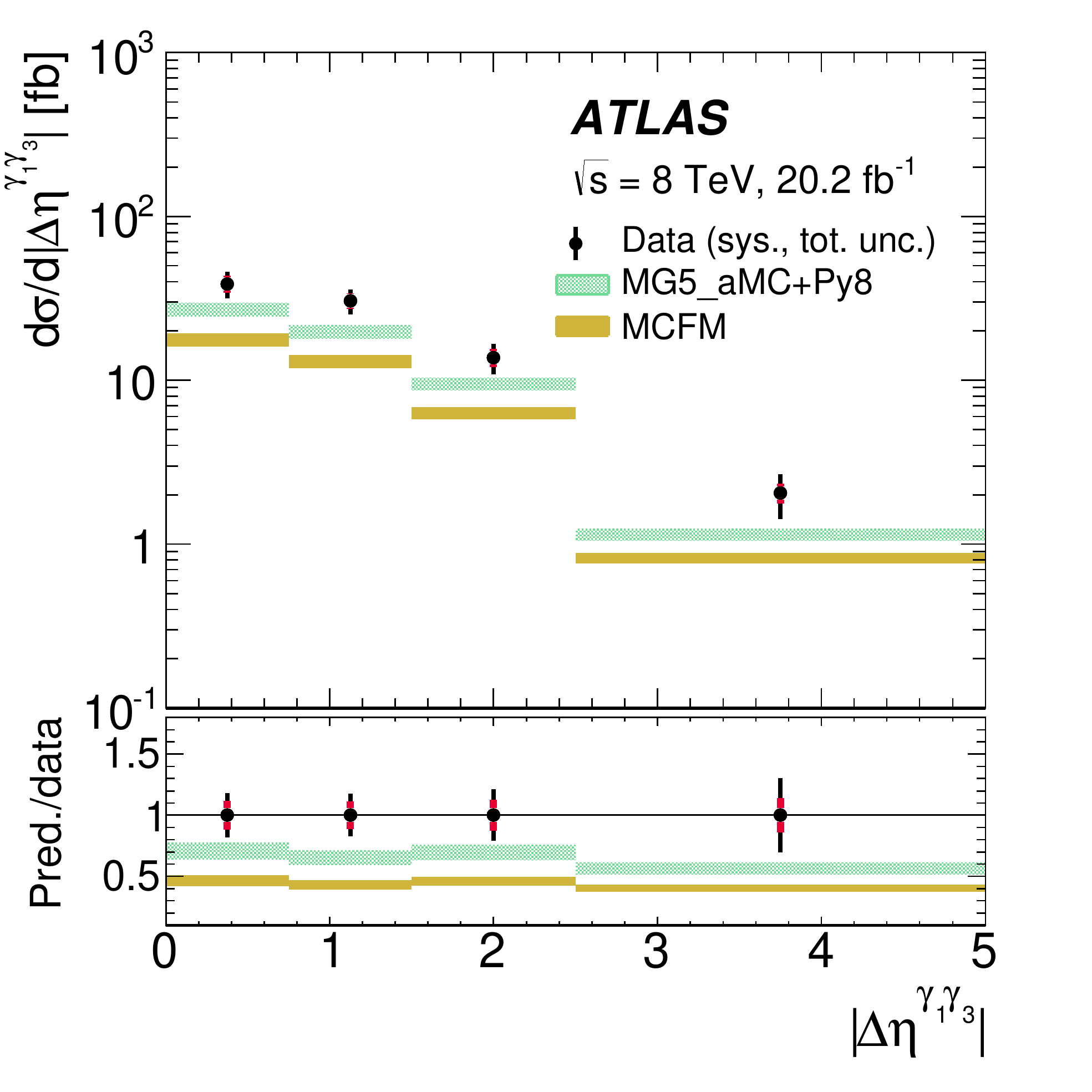}}
  \hspace{0.0cm}\subfloat[]{\includegraphics[scale=0.305]{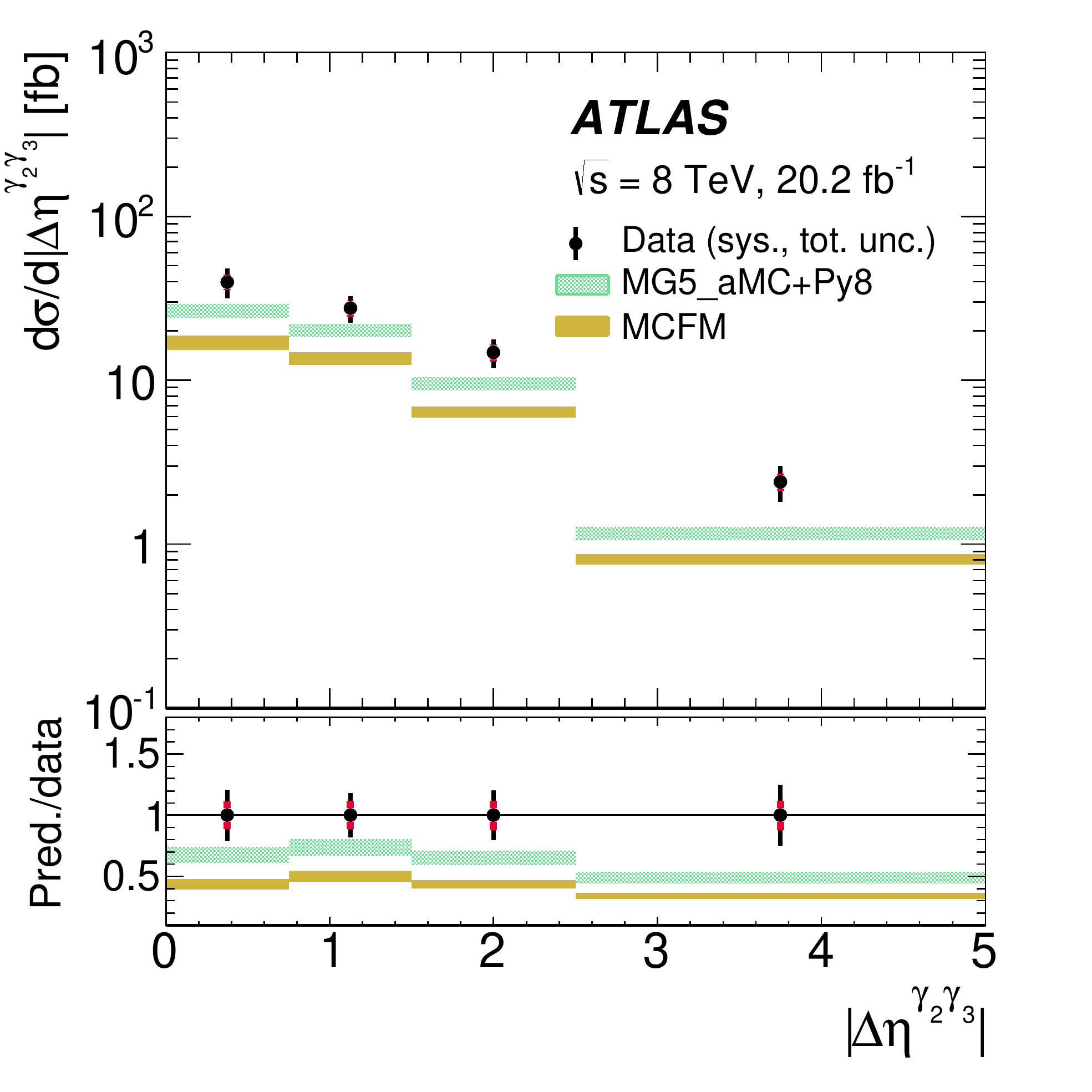}}
  \caption{Measured differential cross sections for the production of three isolated photons (dots) as functions of (a) $\Delta\phi^{\gamma_{1}\gamma_{2}}$, 
  (b) $\Delta\phi^{\gamma_{1}\gamma_{3}}$, (c) $\Delta\phi^{\gamma_{2}\gamma_{3}}$, (d) $|\Delta\eta^{\gamma_{1}\gamma_{2}}|$, 
  (e) $|\Delta\eta^{\gamma_{1}\gamma_{3}}|$ and (f) $|\Delta\eta^{\gamma_{2}\gamma_{3}}|$. The NLO QCD calculations from {\textsc MCFM} and {{\textsc MadGraph5}\_aMC@NLO} are also shown. The thickness of each theoretical prediction corresponds to the theoretical uncertainty. The bottom part of each figure shows the ratios of predicted and measured differential cross sections. The red inner (black outer) error bars represent the systematic uncertainties (the statistical and systematic uncertainties added in quadrature). For some of the data points, the inner error bars are smaller than the marker size and thus not visible.}
  \label{fig:result_mcfm3}
\end{center}
\end{figure}

Figure~\ref{fig:result_mcfm3} shows the three-isolated-photons differential cross sections as functions of $\Delta\phi^{\gamma_{1}\gamma_{2}}$, $\Delta\phi^{\gamma_{1}\gamma_{3}}$, $\Delta\phi^{\gamma_{2}\gamma_{3}}$, $|\Delta\eta^{\gamma_{1}\gamma_{2}}|$, $|\Delta\eta^{\gamma_{1}\gamma_{3}}|$ and $|\Delta\eta^{\gamma_{2}\gamma_{3}}|$. The theoretical calculations underestimate the normalisation of the measurements. This is due to the fact that these distributions are mainly populated by low-$E_{\text T}^{\gamma}$ photons. Both NLO QCD calculations give an adequate description of the shape of the differential cross sections as functions of $|\Delta\eta^{\gamma_{1}\gamma_{2}}|$, $|\Delta\eta^{\gamma_{1}\gamma_{3}}|$ and $|\Delta\eta^{\gamma_{2}\gamma_{3}}|$.  A quantitative comparison of the NLO QCD predictions to the measurements as functions of $\Delta\phi^{\gamma_{1}\gamma_{2}}$, $\Delta\phi^{\gamma_{1}\gamma_{3}}$ and $\Delta\phi^{\gamma_{2}\gamma_{3}}$ is performed with a $\chi^{2}$ fit to the cross-section normalisation including both statistical and systematic uncertainties. This tests the description of the shape of the differential cross sections. The total systematic uncertainty is considered to be fully correlated across bins and is included in the $\chi^{2}$ definition using nuisance parameters. After the $\chi^{2}$ minimisation, scale factors equal to $\approx$ 1.6 ({{\textsc MadGraph5}\_aMC@NLO}) and $\approx$ 2.3 ({\textsc MCFM}) are found for each angular distribution independently. Both theoretical predictions give an adequate description of the shape of $d\sigma/d\Delta\phi^{\gamma_{2}\gamma_{3}}$ ($\chi^{2}/{\text{ndof}}$~=~6/5~and~7/5 for {{\textsc MadGraph5}\_aMC@NLO} and {\textsc MCFM}, respectively, where ndof is the number of degree of freedom). In addition, {{\textsc MadGraph5}\_aMC@NLO} calculations describe adequately the shape of $d\sigma/d\Delta\phi^{\gamma_{1}\gamma_{2}}$ and $d\sigma/d\Delta\phi^{\gamma_{1}\gamma_{3}}$ ($\chi^{2}/{\text{ndof}}$~=~6/5 and ~7/5, respectively) but not {\textsc MCFM} ($\chi^{2}/{\text{ndof}}$~=~13/5 and~14/5, respectively). This shows the importance of the addition of the parton shower to improve the description of the shape of $d\sigma/d\Delta\phi^{\gamma_{1}\gamma_{2}}$ and $d\sigma/d\Delta\phi^{\gamma_{1}\gamma_{3}}$.

\vspace{-0.2cm}

\FloatBarrier
\section{Summary}
\label{sec:summary}

A measurement of the production cross section of three isolated photons in $pp$ collisions at $\sqrt{s}$ = 8~TeV with the ATLAS detector at the LHC is presented using a data set with an integrated luminosity of 20.2~fb$^{-1}$. Differential cross sections as functions of $E_{\text T}^{\gamma_{1}}$, $E_{\text T}^{\gamma_{2}}$, $E_{\text T}^{\gamma_{3}}$, $m^{\gamma_{1}\gamma_{2}}$, $m^{\gamma_{1}\gamma_{3}}$, $m^{\gamma_{2}\gamma_{3}}$, $m^{\gamma\gamma\gamma}$, $\Delta\phi^{\gamma_{1}\gamma_{2}}$, $\Delta\phi^{\gamma_{1}\gamma_{3}}$, $\Delta\phi^{\gamma_{2}\gamma_{3}}$, $|\Delta\eta^{\gamma_{1}\gamma_{2}}|$, $|\Delta\eta^{\gamma_{1}\gamma_{3}}|$, and $|\Delta\eta^{\gamma_{2}\gamma_{3}}|$ are measured for photons with $E_{\text T}^{\gamma_{1}}>$ 27~GeV, $E_{\text T}^{\gamma_{2}}>$ 22~GeV, $E_{\text T}^{\gamma_{3}}>$ 15~GeV, $m^{\gamma\gamma\gamma}>$ 50~GeV, and $|\eta^{\gamma}|~<$~2.37, excluding the region 1.37 $<|\eta^{\gamma}|<$ 1.56. The distance between pairs of photons in the $\eta$--$\phi$ plane is required to be $\Delta R>$ 0.45. The selection of isolated photons is ensured by requiring that the transverse energy in a cone of size $\Delta R$~=~0.4 around the photon is smaller than 10~GeV.

The inclusive fiducial cross section is measured to be $\sigma_{\text{meas}}$ = 72.6~$\pm$~6.5~({\text{stat.}})~$\pm$~9.2~({\text{syst.}})~fb. The NLO QCD calculations underestimate the measured inclusive fiducial cross section by a factor 2.3 for {\textsc MCFM} and 1.6 for {{\textsc MadGraph5}\_aMC@NLO}. Both NLO QCD predictions underestimate the measurements in the low transverse energy and invariant mass regions. The {{\textsc MadGraph5}\_aMC@NLO} predictions give an adequate description  of the measured cross-section distributions for $E_{\text T}^{\gamma_{2}}\gtrsim $ 50 GeV and $E_{T}^{\gamma_{3}}\gtrsim $ 30~GeV and for $m^{\gamma_{1}\gamma_{2}} \gtrsim$~150~GeV, $m^{\gamma_{1}\gamma_{3}} \gtrsim$~75~GeV, $m^{\gamma_{2}\gamma_{3}} \gtrsim$~75~GeV and $m^{\gamma\gamma\gamma} \gtrsim$~150~GeV. Both NLO calculations give an adequate description of the shape of the measured cross section as functions of $|\Delta\eta^{\gamma_{1}\gamma_{2}}|$, $|\Delta\eta^{\gamma_{1}\gamma_{3}}|$ and $|\Delta\eta^{\gamma_{2}\gamma_{3}}|$, whereas they underestimate the normalisation of the measurements. In addition, both theoretical predictions inadequately describe the normalisation of the measurements as functions of $\Delta\phi^{\gamma_{1}\gamma_{2}}$, $\Delta\phi^{\gamma_{1}\gamma_{3}}$ and $\Delta\phi^{\gamma_{2}\gamma_{3}}$. MCFM predictions give an adequate description of the shape of $d\sigma/d\Delta\phi^{\gamma_{2}\gamma_{3}}$ and fail to describe the shape of $d\sigma/d\Delta\phi^{\gamma_{1}\gamma_{2}}$ and $d\sigma/d\Delta\phi^{\gamma_{1}\gamma_{3}}$, whereas {{\textsc MadGraph5}\_aMC@NLO} predictions give an adequate description of the shape of the measured cross sections as functions of all three angular variables. The measurements provide a test of pQCD for the description of the dynamics of triphoton production and indicate the need for improved modelling of this process in MC models.

\section*{Acknowledgements}


We thank CERN for the very successful operation of the LHC, as well as the
support staff from our institutions without whom ATLAS could not be
operated efficiently.

We acknowledge the support of ANPCyT, Argentina; YerPhI, Armenia; ARC, Australia; BMWFW and FWF, Austria; ANAS, Azerbaijan; SSTC, Belarus; CNPq and FAPESP, Brazil; NSERC, NRC and CFI, Canada; CERN; CONICYT, Chile; CAS, MOST and NSFC, China; COLCIENCIAS, Colombia; MSMT CR, MPO CR and VSC CR, Czech Republic; DNRF and DNSRC, Denmark; IN2P3-CNRS, CEA-DRF/IRFU, France; SRNSFG, Georgia; BMBF, HGF, and MPG, Germany; GSRT, Greece; RGC, Hong Kong SAR, China; ISF, I-CORE and Benoziyo Center, Israel; INFN, Italy; MEXT and JSPS, Japan; CNRST, Morocco; NWO, Netherlands; RCN, Norway; MNiSW and NCN, Poland; FCT, Portugal; MNE/IFA, Romania; MES of Russia and NRC KI, Russian Federation; JINR; MESTD, Serbia; MSSR, Slovakia; ARRS and MIZ\v{S}, Slovenia; DST/NRF, South Africa; MINECO, Spain; SRC and Wallenberg Foundation, Sweden; SERI, SNSF and Cantons of Bern and Geneva, Switzerland; MOST, Taiwan; TAEK, Turkey; STFC, United Kingdom; DOE and NSF, United States of America. In addition, individual groups and members have received support from BCKDF, the Canada Council, CANARIE, CRC, Compute Canada, FQRNT, and the Ontario Innovation Trust, Canada; EPLANET, ERC, ERDF, FP7, Horizon 2020 and Marie Sk{\l}odowska-Curie Actions, European Union; Investissements d'Avenir Labex and Idex, ANR, R{\'e}gion Auvergne and Fondation Partager le Savoir, France; DFG and AvH Foundation, Germany; Herakleitos, Thales and Aristeia programmes co-financed by EU-ESF and the Greek NSRF; BSF, GIF and Minerva, Israel; BRF, Norway; CERCA Programme Generalitat de Catalunya, Generalitat Valenciana, Spain; the Royal Society and Leverhulme Trust, United Kingdom.

The crucial computing support from all WLCG partners is acknowledged gratefully, in particular from CERN, the ATLAS Tier-1 facilities at TRIUMF (Canada), NDGF (Denmark, Norway, Sweden), CC-IN2P3 (France), KIT/GridKA (Germany), INFN-CNAF (Italy), NL-T1 (Netherlands), PIC (Spain), ASGC (Taiwan), RAL (UK) and BNL (USA), the Tier-2 facilities worldwide and large non-WLCG resource providers. Major contributors of computing resources are listed in Ref.~\cite{ATL-GEN-PUB-2016-002}.

\printbibliography

\clearpage 
\begin{flushleft}
{\Large The ATLAS Collaboration}

\bigskip

M.~Aaboud$^\textrm{\scriptsize 137d}$,
G.~Aad$^\textrm{\scriptsize 88}$,
B.~Abbott$^\textrm{\scriptsize 115}$,
O.~Abdinov$^\textrm{\scriptsize 12}$$^{,*}$,
B.~Abeloos$^\textrm{\scriptsize 119}$,
S.H.~Abidi$^\textrm{\scriptsize 161}$,
O.S.~AbouZeid$^\textrm{\scriptsize 139}$,
N.L.~Abraham$^\textrm{\scriptsize 151}$,
H.~Abramowicz$^\textrm{\scriptsize 155}$,
H.~Abreu$^\textrm{\scriptsize 154}$
% ...
\bigskip
\\
% ...
$^{12}$ Institute of Physics, Azerbaijan Academy of Sciences, Baku, Azerbaijan\\
% ...
$^{88}$ CPPM, Aix-Marseille Universit{\'e} and CNRS/IN2P3, Marseille, France\\
% ...
$^{115}$ Homer L. Dodge Department of Physics and Astronomy, University of Oklahoma, Norman OK, United States of America\\
% ...
$^{119}$ LAL, Univ. Paris-Sud, CNRS/IN2P3, Universit{\'e} Paris-Saclay, Orsay, France\\
% ...
$^{137}$ $^{(a)}$ Facult{\'e} des Sciences Ain Chock, R{\'e}seau Universitaire de Physique des Hautes Energies - Universit{\'e} Hassan II, Casablanca; $^{(b)}$ Centre National de l'Energie des Sciences Techniques Nucleaires, Rabat; $^{(c)}$ Facult{\'e} des Sciences Semlalia, Universit{\'e} Cadi Ayyad, LPHEA-Marrakech; $^{(d)}$ Facult{\'e} des Sciences, Universit{\'e} Mohamed Premier and LPTPM, Oujda; $^{(e)}$ Facult{\'e} des sciences, Universit{\'e} Mohammed V, Rabat, Morocco\\
% ...
$^{139}$ Santa Cruz Institute for Particle Physics, University of California Santa Cruz, Santa Cruz CA, United States of America\\
% ...
$^{151}$ Department of Physics and Astronomy, University of Sussex, Brighton, United Kingdom\\
% ...
$^{154}$ Department of Physics, Technion: Israel Institute of Technology, Haifa, Israel\\
$^{155}$ Raymond and Beverly Sackler School of Physics and Astronomy, Tel Aviv University, Tel Aviv, Israel\\
% ...
$^{161}$ Department of Physics, University of Toronto, Toronto ON, Canada\\
% ...
$^{a}$ Also at Department of Physics, King's College London, London, United Kingdom\\
$^{b}$ Also at Institute of Physics, Azerbaijan Academy of Sciences, Baku, Azerbaijan\\
$^{c}$ Also at Novosibirsk State University, Novosibirsk, Russia\\
$^{d}$ Also at TRIUMF, Vancouver BC, Canada\\
$^{e}$ Also at Department of Physics {\&} Astronomy, University of Louisville, Louisville, KY, United States of America\\
% ...
$^{*}$ Deceased
\end{flushleft}


\end{document}